\documentclass[preprint,floatfix]{elsarticle}

\usepackage{graphicx}
\usepackage{amsmath}
\usepackage{amssymb}
\usepackage{subcaption}
\usepackage{xcolor}

\begin{document}

\title{Discrete $\mathcal{PT}$-symmetric solitons on branched lattices: \\
Vertex states, stability, and vortices}

\author[*]{M. Akramov\corref{cor1}}
\ead{mashrabresearcher@gmail.com}
\author[*]{A. Fayzullaev}
\author[*]{O. Tojakhmadova}
\author[*]{T. Akhmadjanov}

\cortext[cor1]{Corresponding author}
\affiliation[*]{organization={National University of Uzbekistan},
addressline={Universitet Str. 4},
postcode={100174},
city={Tashkent},
country={Uzbekistan}}

\begin{abstract}
Branching fundamentally modifies nonlinear localization by introducing vertex degrees of freedom absent in one-dimensional lattices. We investigate localized states in a $\mathcal{PT}$-symmetric four-edge star graph described by the discrete nonlinear Schr\"odinger equation. We identify a discrete vortex state with a fixed topological charge, supported by the branching geometry. Constructed by continuation from the anticontinuum limit, the vortex evolves into a current-carrying state in the presence of balanced gain and loss and exhibits stability windows absent for the conventional branches. A possible electric-circuit realization of the proposed model is discussed.
\end{abstract}

\maketitle

\section{Introduction}\label{section1}

Parity-time ($\mathcal{PT}$) symmetric systems have attracted considerable attention in non-Hermitian physics since Hamiltonians respecting $\mathcal{PT}$ symmetry were shown to exhibit entirely real spectra below a critical threshold. Originally introduced in optics, $\mathcal{PT}$ symmetry has since been demonstrated experimentally in coupled waveguides \cite{guo_observation_2009,ruter_observation_2010}, photonic lattices \cite{regensburger_paritytime_2012}, resonators \cite{peng_paritytime-symmetric_2014,chang_paritytime_2014}, and electric circuits \cite{wang_observation_2020,tabeu_imaginary_2019}.

The interplay of balanced gain-loss and nonlinearity supports localized states in $\mathcal{PT}$-symmetric lattices despite their non-conservative nature \cite{konotop_nonlinear_2016}. These nonlinear modes have been observed experimentally in optical lattices \cite{wimmer_observation_2015}. Most studies, however, have focused on one-dimensional lattices, while the role of branching and network topology in nonlinear localization remains much less understood.

The existence and stability of discrete solitons in one-dimensional $\mathcal{PT}$-symmetric lattices have been extensively investigated. Beginning with the continuation from the anticontinuum limit in infinite $\mathcal{PT}$-symmetric chains \cite{Konotop_2012}, this framework has been extended to finite lattices, defect configurations, and more complex $\mathcal{PT}$-symmetric building blocks, together with their nonlinear dynamics and stability properties \cite{kevrekidis_nonlinear_2013b,li_span_2011,li_nonlinear_2012,duanmu_linear_2013,kevrekidis_nonlinear_2013,zezyulin_stationary_2013,zezyulin_nonlinear_2012,pelinovsky_nonlinear_2014,suchkov_solitons_2011}.

Many physical systems, however, possess branched rather than one-dimensional geometries. Such networks are naturally modeled by quantum or metric graphs with coupling conditions imposed at the vertices \cite{exner_quantum_1988,kostrykin_kirchhoffs_1999,kottos_periodic_1999,berkolaiko_introduction_2013}. Although nonlinear waves on branched structures and $\mathcal{PT}$-symmetric graph models have received increasing attention \cite{sobirov_integrable_2010,nakamura_transport_2011,akramov_nonlocal_2022,akramov_discrete_2023,akramov_pt-symmetric_2024}, the branching introduces a fundamentally different coupling mechanism, requiring the existence and stability analysis to be reformulated beyond the one-dimensional setting.

In the anticontinuum limit, where inter-site coupling vanishes and the lattice decouples into isolated sites, the stationary equations at the branching vertex admit two natural localized configurations: symmetric and antisymmetric states. These, however, represent only special solutions of the full vertex problem. More generally, the vertex equations allow unequal edge amplitudes whose phases wind around the branching point, giving rise to a nonzero discrete winding number \cite{malomed_discrete_2001,neshev_observation_2004,leykam_discrete_2013}. Such phase-winding states have no analogue in a simple one-dimensional chain, where no closed path exists. Motivated by this observation, we study the existence and stability of vertex-localized $\mathcal{PT}$-symmetric vortex solitons on a branched lattice.

In this work, we investigate localized states in a $\mathcal{PT}$-symmetric four-edge star graph described by the discrete nonlinear Schrödinger equation. We first determine the linear $\mathcal{PT}$-symmetric spectrum and identify the parameter regime of unbroken $\mathcal{PT}$ symmetry. We then analyze the existence and stability of the symmetric and antisymmetric vertex states and demonstrate the emergence of a distinct vortex branch characterized by a nonzero discrete winding number. Finally, we compare the stability properties of the vortex and non-vortex states, highlighting the role of the branching topology in supporting phase-winding localized modes.

We show that the vortex soliton carries a nonzero discrete winding number and evolves into a current-carrying state in the presence of balanced gain and loss. While its existence follows the general conditions for vertex-localized modes, its stability differs qualitatively from that of the conventional symmetric and antisymmetric solutions, exhibiting finite stability windows enabled by the branching topology.

These results demonstrate that the interplay of branching, nonlinearity, and $\mathcal{PT}$ symmetry supports a distinct class of localized phase-winding states that is absent in one-dimensional $\mathcal{PT}$ lattices. The remainder of the paper is organized as follows. Section~\ref{section2} introduces the $\mathcal{PT}$-symmetric discrete chain and its Gaussian gain-loss extension. Section~\ref{section3} develops the star-graph model and analyzes the symmetric and antisymmetric vertex states. Section~\ref{sec:vortex} constructs and investigates the vortex soliton. Section~\ref{sketch} discusses a possible electric-circuit realization. Finally, Section~\ref{sec:conclusion} gives concluding remarks and future perspectives.

\section{$\mathcal{PT}$-symmetric discrete nonlinear Schr\"odinger equation on a chain}
\label{section2}

\subsection{Konotop model}
In this subsection, we briefly recall the $\mathcal{PT}$-symmetric discrete
nonlinear Schr\"odinger (NLDS) equation, following \cite{Konotop_2012}:
\begin{equation}\label{NLDS}
    i \frac{dq_n}{dt} + c_n (q_{n+1} - q_n) + c_{n+1} (q_{n-1} - q_n) - g \lvert q_n \rvert^2 q_n
    + i(-1)^{n+1} \gamma q_n = 0,
\end{equation}
where the positive constants $c_n = \kappa_0$ for $n=2p$ and $c_n = \kappa_1$
for $n=2p+1$ describe the coupling coefficients between neighboring sites,
and $\gamma > 0$ characterizes the strength of the gain and loss at
alternating sites.

In the linear case, i.e., $g=0$, identifying solutions of the linear problem
in the Floquet--Bloch form $(q_{2n}, q_{2n+1}) = (u,v)\, e^{ikn - i(\kappa_0 + \kappa_1 + \mu)t}$,
one obtains the $\mathcal{PT}$-unbroken condition
\begin{equation}\label{threshold}
    |\kappa_0 - \kappa_1|\geq \gamma,
\end{equation}
equivalently, $\mu^2$ lies in the interval
\begin{equation}
    (\kappa_0 - \kappa_1)^2 - \gamma^2 < \mu^2 < (\kappa_0 + \kappa_1)^2 - \gamma^2.
\end{equation}
Condition \eqref{threshold} governs the reality of the eigenvalue $\mu$: if it
is not satisfied, $\mu$ acquires a nonzero imaginary part and the
$\mathcal{PT}$ symmetry is broken.

In what follows localized (soliton) solutions can be obtained via the
envelope ansatz for the gain amplitude $u_n$ and the loss amplitude $v_n$,
given by
\begin{equation}\label{separation}
    \begin{pmatrix}
        q_{2n}(t) \\
        q_{2n+1}(t)        
    \end{pmatrix}
     = 
     \begin{pmatrix}
        u_n \\
        v_n        
    \end{pmatrix}
    e^{-i(\kappa_0 + \kappa_1 + \mu)t},
\end{equation}
where $u_n$ and $v_n$ satisfy the asymptotic boundary conditions
$u_n, v_n \to 0$ as $n\to\pm \infty$.

The NLDS equation \eqref{NLDS} can then be rewritten as a system of coupled
nonlinear algebraic equations,
\begin{equation}
    \begin{cases} \label{eq:algebraic}
        (\mu - i\gamma)u_n + \kappa_0 v_n + \kappa_1 v_{n-1} = g|u_n|^2 u_n,\\
        (\mu + i\gamma)v_n + \kappa_0 u_n + \kappa_1 u_{n+1} = g|v_n|^2 v_n.
    \end{cases}
\end{equation}
Here the $\mathcal{PT}$-symmetry condition is written in the form 
\begin{equation}
    u_n=\bar v_{-n}, \quad
    v_n=\bar u_{-n},
\end{equation}
where $\bar{}$ means the complex conjugate.
Nontrivial, exponentially localized solutions of \eqref{eq:algebraic}
satisfying \eqref{separation} were constructed in \cite{Konotop_2012} by
analytical continuation from the anticontinuum limit $\kappa_1 = 0$, where
the lattice decouples into a set of isolated $\mathcal{PT}$-symmetric dimers.
Two families of stationary solitons, bifurcate from
the single-dimer states and persist, under appropriate conditions, to
$\kappa_1 > 0$. These families and their linear stability are the basis for
the analysis in the next sections. Using this Konotop model, we consider, in the following subsection, its generalization to a Gaussian
gain--loss profile.

\subsection{Gaussian gain-loss generalization}

Now we generalize the $\mathcal{PT}$-symmetric discrete nonlinear
Schr\"odinger (NLDS) equation by replacing the constant gain-loss
coefficient with a Gaussian profile,
\begin{equation}\label{eq:NLDS-V}
    i \frac{dq_n}{dt} + c_n (q_{n+1} - q_n) + c_{n+1} (q_{n-1} - q_n) - g \lvert q_n \rvert^2 q_n
    + i(-1)^{n+1} V_n q_n = 0,
\end{equation}
where
\begin{equation}\label{Gaussian}
    \begin{pmatrix}
        V_{2n} \\ V_{2n+1}
    \end{pmatrix} = \gamma_n, \quad
    \gamma_n = \gamma_0 \exp\bigg(-\frac{n^2} {\sigma^2}\bigg).
\end{equation}
Here $\sigma$ defines the width of the profile. 

We recover Konotop's model in the limit $\sigma\to \pm\infty$, while the
gain-loss profile vanishes as $\sigma\to0$ and the system reduces to
the conservative lattice (up to a residual dimer at $n=0$).

Using the Eq.~\eqref{separation}, we obtain the reduced
system of equations
\begin{equation}
    \begin{cases} \label{uv}
        (\mu - i\gamma_n)u_n + \kappa_0 v_n + \kappa_1 v_{n-1} = g|u_n|^2 u_n,\\
        (\mu + i\gamma_n)v_n + \kappa_0 u_n + \kappa_1 u_{n+1} = g|v_n|^2 v_n.
    \end{cases}
\end{equation}

In the linear regime $g=0$, equation \eqref{uv} reduces to the eigenvalue
problem
\begin{equation}\label{eig1}
    \begin{cases}
         i\gamma_n u_n - \kappa_0 v_n - \kappa_1 v_{n-1} = \mu u_n,\\
        -i\gamma_n v_n - \kappa_0 u_n - \kappa_1 u_{n+1} = \mu v_n.
    \end{cases}
\end{equation}

The eigenvalue problem in \eqref{eig1} can be rewritten in matrix form as
\begin{equation}
    H\omega=\mu\omega,
\end{equation}
where 
\[
\omega=\big(\ldots,\omega_{-1},\omega_0,\omega_1,\ldots\big)^T,
\qquad
\omega_n=(u_n,v_n)^T,
\]
and $H$ is an infinite block-tridiagonal operator of the form
\begin{equation}\label{H1}
H=
\begin{pmatrix}
\ddots & \ddots & \ddots & & & \\
\ddots & D_{-1} & T_{+} & 0 & & \\
\ddots & T_{-} & D_{0} & T_{+} & \ddots & \\
& 0 & T_{-} & D_{1} & \ddots & \\
& & \ddots & \ddots & \ddots &
\end{pmatrix},
\end{equation}
with the $2\times 2$ blocks
\begin{equation}\label{DT}
D_n=
\begin{pmatrix}
i\gamma_n & -\kappa_0\\
-\kappa_0 & -i\gamma_n
\end{pmatrix},
\qquad
T_{+}=
\begin{pmatrix}
0 & 0\\
-\kappa_1 & 0
\end{pmatrix},
\qquad
T_{-}=
\begin{pmatrix}
0 & -\kappa_1\\
0 & 0
\end{pmatrix}.
\end{equation}

Equivalently, the eigenvalue problem can be written in the cell form
\begin{equation}\label{bulk1}
T_{-}\omega_{n-1}+D_n\omega_n+T_{+}\omega_{n+1}=\mu\,\omega_n,
\qquad n\in\mathbb{Z}.
\end{equation}

\begin{figure}[t!]
    \centering
    \includegraphics[width=0.49\linewidth]{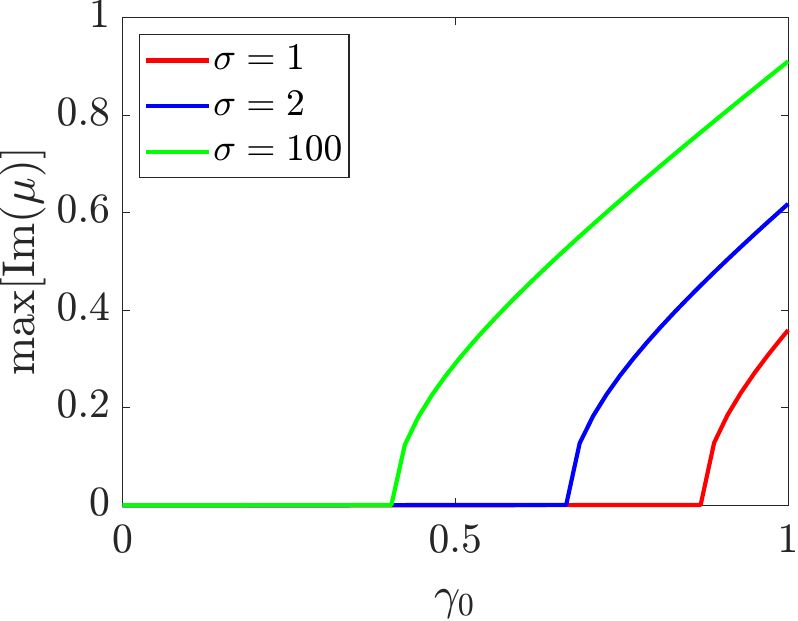}
    \caption{Maximal imaginary part of the eigenvalue $\mu$ of
    the Hamiltonian $H$ in \eqref{H1} versus $\gamma_0$, for three values of
    the localization parameter $\sigma$, other parameters are chosen as $\kappa_0=1, \; \kappa_1=0.6$. 
    }
    \label{fig:im_mu}
\end{figure}

We numerically compute the spectrum of the Hamiltonian $H$ in \eqref{H1} and
plot the dependence of the maximal imaginary part of the eigenvalue $\mu$ on
the parameter $\gamma_0$ (see Fig.~\ref{fig:im_mu}). As the
gain-loss profile becomes narrower with decreasing $\sigma$, the
$\mathcal{PT}$ phase transition threshold shifts to higher values of
$\gamma_0$. This behavior is expected, since increasing the localization of
the gain-loss profile reduces the number of lattice sites participating in
the non-Hermitian interaction, thus requiring larger values of $\gamma_0$ to
induce $\mathcal{PT}$-symmetry breaking.

Having established the linear spectral properties of the lattice with a
Gaussian gain-loss profile, we next turn, in the following subsection, to
the nonlinear regime $g\neq0$.

\subsection{Anticontinuum limit and existence of solitons}\label{subsec:existence}

We establish the existence of localized soliton solutions of the nonlinear
system \eqref{uv} by analytical continuation from the anticontinuum limit,
following the approach of \cite{Konotop_2012}.

In the anticontinuum limit $\kappa_1 = 0$, the lattice \eqref{uv} decouples
into a sequence of independent dimers indexed by $n\in\mathbb{Z}$, each
detuned by the local gain-loss coefficient $\gamma_n$:
\begin{equation}\label{eq:dimer-n}
    (\mu - i\gamma_n)u_n + \kappa_0 v_n = g|u_n|^2 u_n, \qquad
    (\mu + i\gamma_n)v_n + \kappa_0 u_n = g|v_n|^2 v_n.
\end{equation}
We consider the simplest case in which only the central dimer, located at
$n=0$ where the gain-loss coefficient attains its maximal value
$\gamma_0$, is excited, i.e., $\omega_n = 0$ for $n\neq 0$. Imposing the
$\mathcal{PT}$-symmetry reduction $u_0 = \bar v_0 =: p$, Eq.~\eqref{eq:dimer-n} reduces to a single complex equation,
\begin{equation}\label{eq:central-dimer}
    (\mu - i\gamma_0)p + \kappa_0 \bar p = g|p|^2 p,
\end{equation}
which is identical in structure to the single-dimer equation of
\cite{Konotop_2012} (with $\kappa_0 = 1$ therein). Writing $p = A e^{i\phi}$
and separating real and imaginary parts gives two exact branches of solutions,
\begin{equation*}
    p_0^{\pm} = A_{\pm} e^{i\phi_{\pm}}, \;
    A_{\pm}^2 = \frac{\mu \pm \sqrt{\kappa_0^2 - \gamma_0^2}}{g}, \;
    \sin(2\phi_\pm) = -\frac{\gamma_0}{\kappa_0}, \; 
    \cos(2\phi_{\pm}) = \pm \frac{\sqrt{\kappa_0^2-\gamma_0^2}}{\kappa_0}
\end{equation*}
which exist provided $\gamma_0 < \kappa_0$ (so that $\cos(2\phi_\pm)$ is
real) and $\mu > \mu_\pm$, where $\mu_\pm = \mp\sqrt{\kappa_0^2-\gamma_0^2}$.
These solution branches are exactly the same as those in~\cite{Konotop_2012}, except that $\kappa_0$ is replaced by 1.

To continue these solutions from $\kappa_1=0$ to $\kappa_1>0$, we impose the
global $\mathcal{PT}$-symmetry reduction $u_n = \bar v_{-n}$, $v_n = \bar
u_{-n}$ and apply the implicit function theorem, invoking the invertibility
of the limiting Jacobian operator by the same argument as in
\cite{Konotop_2012}. We thus obtain two branches of discrete solitons,
denoted $(+)$ and $(-)$, which bifurcate from the
central-dimer states and persist, for sufficiently small $\kappa_1$, on the
chain with the Gaussian gain-loss profile.

\subsection{Linear stability of discrete $\mathcal{PT}$ solitons}
Let $\omega_n = (u_n,v_n)^T$ denote a stationary solution of the system
\eqref{uv}. Following the same approach as in~\cite{Konotop_2012}, we
study the linear stability of such discrete $\mathcal{PT}$-symmetric
solitons by introducing perturbations to $\omega_n$, of the form
\begin{equation}
    \begin{pmatrix}
        q_{2n}(t) \\
        q_{2n+1}(t)        
    \end{pmatrix}
     = \left(
     \omega_n+
    e^{\lambda t}\psi_{n}+
     e^{\bar\lambda t}\bar\varphi_{n}
    \right)
    e^{-i(\kappa_0 + \kappa_1 + \mu)t},
\end{equation}
where $\psi_{n}=(\psi_{u,n},\psi_{v,n})^T$, $\varphi_{n}=(\varphi_{u,n},\varphi_{v,n})^T$, 
and $|\psi_n|, \, |\varphi_n| \ll |\omega_n|$; the complex growth rate
$\lambda$ determines stability, with the soliton being linearly stable if
and only if $\mathrm{Re}(\lambda) \leq 0$ for every eigenvalue of the
resulting problem.

Substituting this ansatz into the NLDS equation \eqref{eq:NLDS-V} and
linearizing with respect to $\psi_n$ and $\varphi_n$ (i.e., neglecting
higher-order terms), we obtain the eigenvalue problem, written in cell form,
as
\begin{equation}\label{stability}
    \begin{pmatrix}
        T_- & 0 \\ 0 & -T_-
    \end{pmatrix}
    \begin{pmatrix}
        \psi_{n-1} \\ \varphi_{n-1}
    \end{pmatrix}+
    \begin{pmatrix}
         L_n & G_n \\ -\bar G_n & -\bar L_n
    \end{pmatrix}
    \begin{pmatrix}
         \psi_n \\ \varphi_n
    \end{pmatrix}+
    \begin{pmatrix}
        T_+ & 0 \\ 0 & -T_+
    \end{pmatrix}
    \begin{pmatrix}
        \psi_{n+1} \\ \varphi_{n+1}
    \end{pmatrix} = 
    i\lambda \begin{pmatrix}
         \psi_n \\ \varphi_n
    \end{pmatrix},
\end{equation}
where
$L_n = [2g \, (|\omega_n|^2) - \mu] I_2 +D_n$,
$G_n = g \, (\omega_n^2) I_2$,
$I_2$ is the $2\times 2$ identity matrix, and $T_{\pm}$, $D_n$ are given in
\eqref{DT}.

\begin{figure}[t!]
    \centering
    \includegraphics[width=0.99\linewidth]{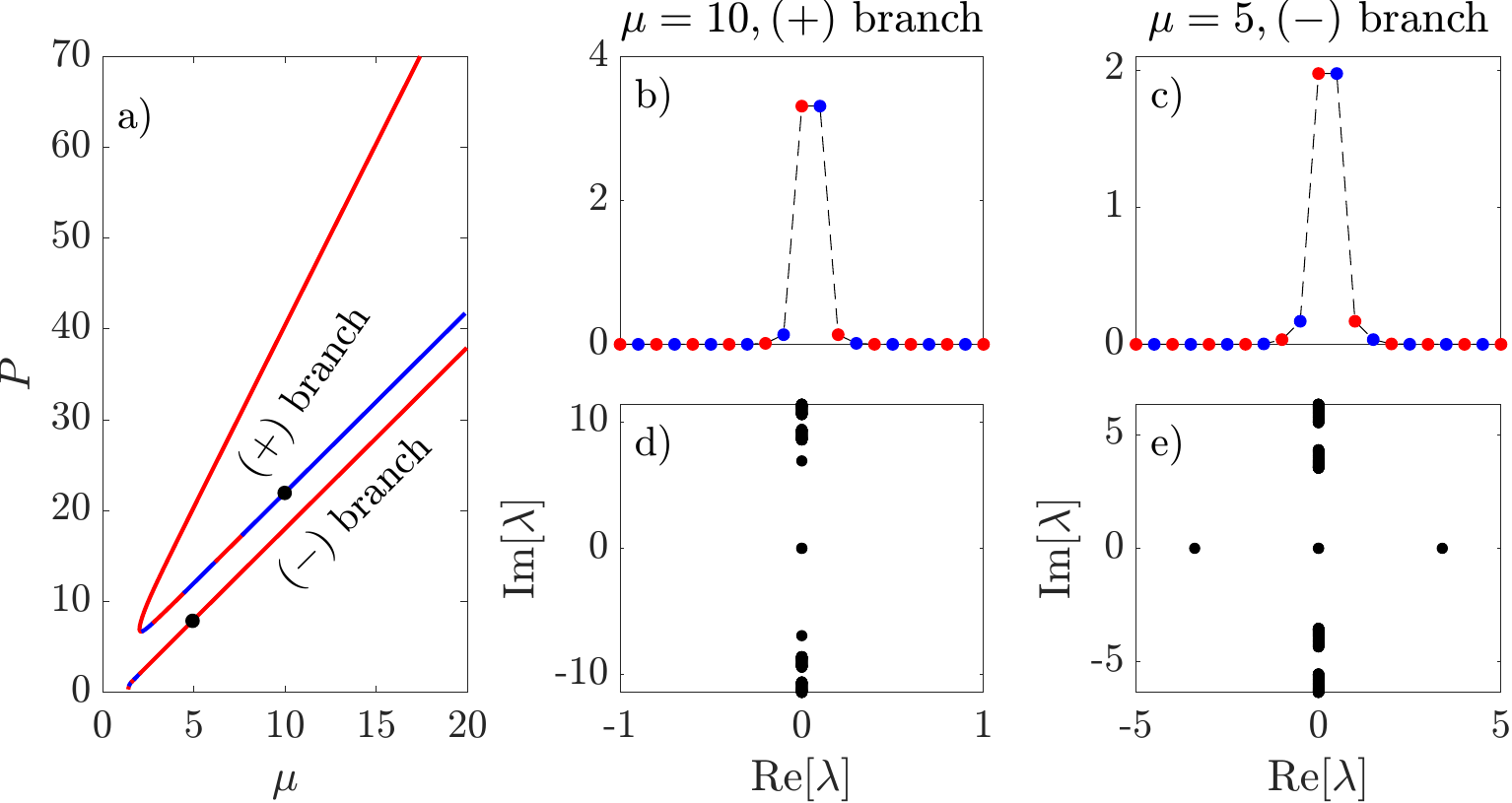}
    \caption{Numerical continuation of discrete $\mathcal{PT}$-symmetric
    solitons bifurcating from the anticontinuum limit for the parameters $\kappa_0=1, \; \kappa_1=0.4, \; \gamma_0=0.1, \; \sigma=2.5, \; g=1$. a) total
    power $P=\sum_n(|u_n|^2+|v_n|^2)$ versus $\mu$ for the two soliton
    branches, blue for the stable and red for the unstable solitons; black dots mark
    the representative points examined in the middle and right columns.
    b) and c) soliton amplitude profiles
    $|\omega_n|^2$, with red and blue dots denoting the gain and loss
    sites, respectively, and d) and e) the corresponding linear stability spectrum
    $\mathrm{Im}[\lambda]$ versus $\mathrm{Re}[\lambda]$.}
    \label{fig:soliton_line}
\end{figure}

Numerical continuation of stationary solutions is performed using a pseudo-arclength continuation scheme, which allows tracing solution branches through fold points (see Figs.~\ref{fig:soliton_line}, \ref{fig:sol_star_full}, \ref{fig:vortex_soliton}). At each step the resulting nonlinear algebraic system is solved by Newton-based trust-region method. This procedure is used for all continuation results presented in this paper.

All numerical results reported in this paper are
obtained on a finite lattice, constructed by truncating each semi-infinite
branch after $N=50$ dimers measured from the reference point (the
central site $n=0$ for the chain, the vertex for the star graph). We
impose a Dirichlet-type boundary condition, the field is set to
zero beyond the truncation point with no absorbing or open boundary
layer.

We solve the eigenvalue problem \eqref{stability} numerically for the two
soliton branches $(+)$ and $(-)$ that bifurcate from the
single-dimer states in the anticontinuum limit.
Fig.~\ref{fig:soliton_line} shows the resulting power versus $\mu$
diagram together with representative amplitude profiles and stability
spectra. The $(-)$ branch, illustrated here at $\mu=5$, exhibits an
isolated pair of eigenvalues with nonzero real part and is therefore
linearly unstable, whereas the $(+)$ branch, illustrated at
$\mu=10$, has its entire spectrum confined to the imaginary axis and is
linearly stable. This asymmetry between the two branches is consistent
with the behavior reported in Ref.~\cite{Konotop_2012} for the chain of
$\mathcal{PT}$-symmetric dimers.

\section{Solitons on branched $\mathcal{PT}$ lattice}
\label{section3}

\subsection{Branched discrete lattices with 4 edges}\label{subsec:star-linear}

\begin{figure}[t!]
    \centering
    \includegraphics[width=0.5\linewidth]{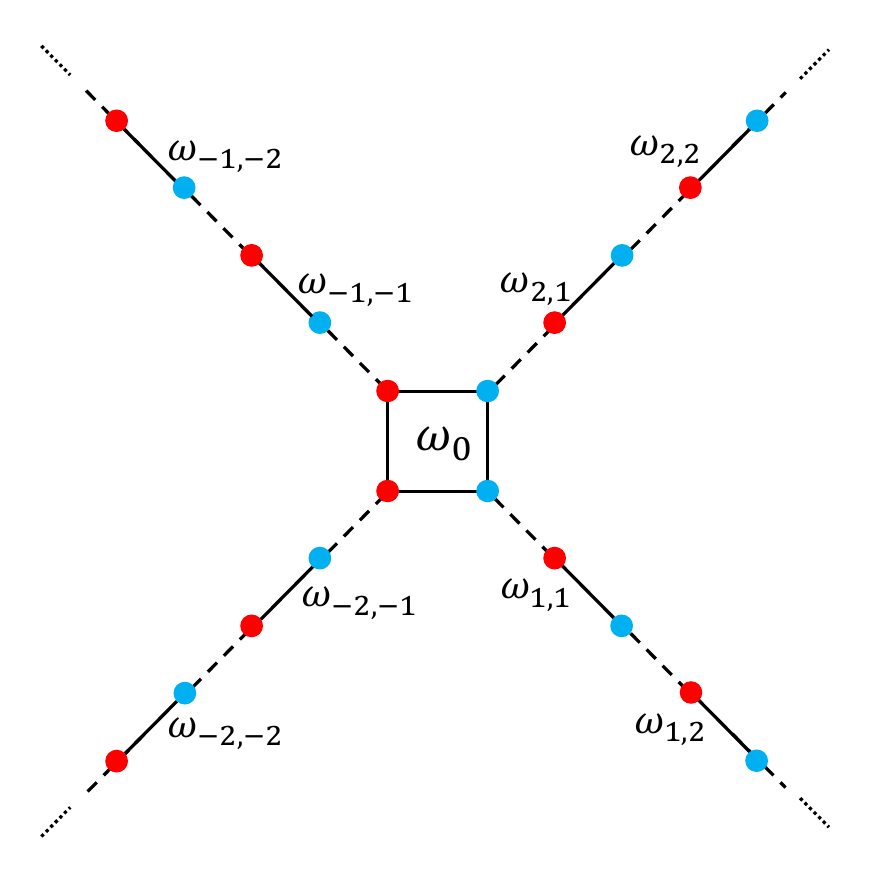}
    \caption{The star graph with four discrete edges. The central quadrimer
    denotes the vertex block $\Omega_0$; red and blue dots mark the gain
    and loss sites, respectively, on each of the four semi-infinite edges.}
    \label{fig:star4}
\end{figure}

We now extend the $\mathcal{PT}$-symmetric NLDS model to a branched
(star-graph) geometry, in which four semi-infinite discrete edges are
joined at a single central vertex (see Fig.~\ref{fig:star4}). Each of the four edges, indexed by 
$j\in\{1,2\}$ (outgoing) and $-j\in\{-1,-2\}$ (incoming), supports
its own copy of the dimer chain, with sites indexed by $n\in\mathbb{Z}$
measured outward from the vertex; edges $j=1,2$ carry outgoing fields and
edges $j=-1,-2$ carry incoming fields, so that the vertex itself is
formed by the four sites $u_{-1,0},v_{1,0},u_{-2,0},v_{2,0}$ where the
edges meet.

At the vertex, the $\mathcal{PT}$-symmetric NLDS equation \eqref{eq:NLDS-V}
couples with other edges through the coupling $\kappa_0$,
\begin{equation}
    \begin{cases} \label{uv1}
    (\mu - i\gamma_0)u_{-j,0} + \kappa_0  (v_{1,0} +v_{2,0}) + \kappa_1 v_{-j,-1} = g|u_{-j,0}|^2 u_{-j,0},\\
    (\mu + i\gamma_0)v_{j,0} + \kappa_0  (u_{-1,0} +u_{-2,0}) + \kappa_1 u_{j,1} = g|v_{j,0}|^2 v_{j,0},
    \end{cases}
\end{equation}
for $j=1,2$. Away from the vertex each edge obeys the same
dimer-chain equation as on the infinite lattice, Eq.~\eqref{uv}, relabeled
by edge index,
\begin{equation}
    \begin{cases} \label{uv3}
    (\mu - i\gamma_{n})u_{\pm j,n} + \kappa_0 v_{\pm j,n} + \kappa_1 v_{\pm j,n-1} = g|u_{\pm j,n}|^2 u_{\pm j,n},\\
    (\mu + i\gamma_{n})v_{\pm j,n} + \kappa_0 u_{\pm j,n} + \kappa_1 u_{\pm j,n+1} = g|v_{\pm j,n}|^2 v_{\pm j,n},
    \end{cases}
\end{equation}
where $n\ge 1$ for edges $+j$ and $n\le -1$ for edges $-j$.

For the four-edge star graph of Fig.~\ref{fig:star4}, we now recast the
linear ($g=0$) equations \eqref{uv1} and \eqref{uv3} as a single
block-tridiagonal eigenvalue problem. Since edges $+j$ (outgoing) and
$-j$ (incoming) are traversed by a common site index $n$, we combine
both edge pairs into a single four-component block $\Omega_n$ defined for
every $n\in\mathbb{Z}$:
\begin{equation}
\Omega_n=
\begin{cases}
\big(u_{1,n},\,v_{1,n},\,u_{2,n},\,v_{2,n}\big)^{T}, & n\ge 1,\\[4pt]
\big(u_{-1,0},\,v_{1,0},\,u_{-2,0},\,v_{2,0}\big)^{T}, & n=0,\\[4pt]
\big(u_{-1,n},\,v_{-1,n},\,u_{-2,n},\,v_{-2,n}\big)^{T}, & n\le -1.
\end{cases}
\label{eq:Omega-n}
\end{equation}
No site is left out and no truncation is introduced. $\Omega_n$ is defined
on the whole line $n\in\mathbb{Z}$, with the branching encoded entirely in
the structure of the coupling matrices below. Away from the vertex, each
edge evolves as two decoupled copies of the chain operator \eqref{DT}.

\begin{figure}[t!]
    \centering
    \includegraphics[width=0.5\linewidth]{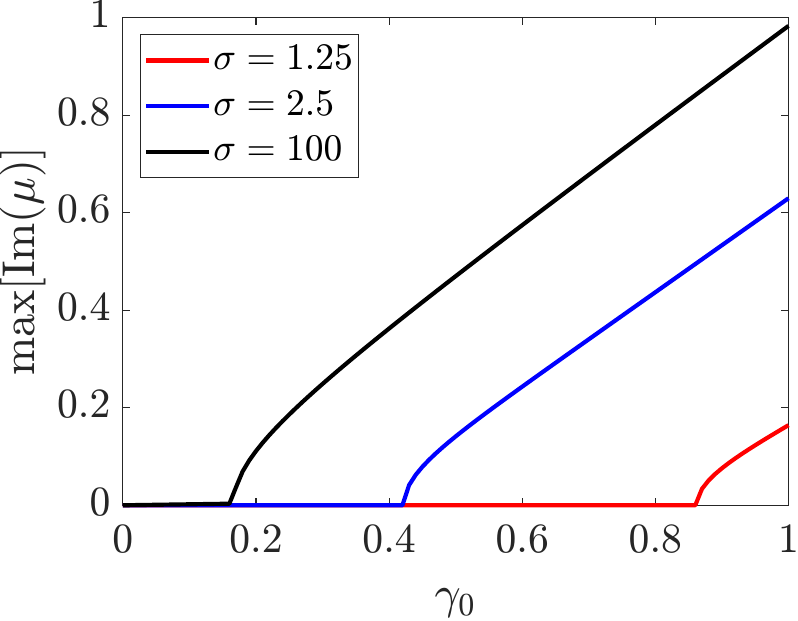}
    \caption{Maximal imaginary part of the eigenvalue $\mu$ of the linear
    vertex operator $\mathcal{H}$ in Eq.~\eqref{eq:H-star} versus $\gamma_0$,
    for three values of the localization $\sigma$. Other parameters are chosen as follows:
    $\kappa_0=1$, $\kappa_1=1.2$. For each
    $\sigma$, $\max[\mathrm{Im}(\mu)]$ vanishes below a critical $\gamma_0$,
    marking the onset of $\mathcal{PT}$-symmetry breaking. This threshold
    increases as the gain-loss profile becomes more localized (smaller
    $\sigma$), consistent with the chain result of Fig.~\ref{fig:im_mu}.}
    \label{fig:max_mu_star}
\end{figure}

We reuse the $2\times2$ blocks $D_n,\,T_+,\,T_-$ already defined in
\eqref{DT}, and introduce new block-diagonal matrices given by
\begin{equation}
\mathcal{D}_n = D_{n}\oplus D_{n},
\qquad
\mathcal{T}_+ = T_+\oplus T_+,
\qquad
\mathcal{T}_- = T_-\oplus T_-,
\qquad |n|\ge 1,
\label{eq:bulk-4x4}
\end{equation}
while the vertex block $\mathcal{D}_0$ additionally couples the two edge
pairs through the off-diagonal coupling $\kappa_0$,
\begin{equation}\label{D0}
    \mathcal{D}_0 = \begin{pmatrix}
        D_0 & -\kappa_0 \sigma_1 \\
        -\kappa_0 \sigma_1 & D_0
    \end{pmatrix},
\end{equation}
where $\sigma_1$ is the Pauli matrix. The bulk eigenvalue equation reads
\begin{equation}
\mathcal{T}_-\,\Omega_{n-1}+\mathcal{D}_n\,\Omega_n+\mathcal{T}_+\,\Omega_{n+1}=\mu\,\Omega_n,
\qquad
n\in \mathbb{Z},
\label{eq:bulk-eq}
\end{equation}
which is identical in structure to Eq.~\eqref{bulk1}. Acting on
$\Omega=(\dots,\Omega_{-1},\Omega_0,\Omega_1,\dots)^T$, the single
block-tridiagonal operator
\begin{equation}
\mathcal{H}=
\begin{pmatrix}
\ddots & \ddots & \ddots & & \\
\ddots & \mathcal{D}_{-1} & \mathcal{T}_+ & 0 & \\
\ddots & \mathcal{T}_- & \mathcal{D}_0 & \mathcal{T}_+ & \ddots \\
 & 0 & \mathcal{T}_- & \mathcal{D}_{1} & \ddots \\
 & & \ddots & \ddots & \ddots
\end{pmatrix},
\qquad \mathcal{H}\,\Omega=\mu\,\Omega,
\label{eq:H-star}
\end{equation}
governs the spectral properties of the star graph.

We numerically diagonalize $\mathcal{H}$ and track the maximal imaginary
part of the eigenvalue $\mu$ as a function of $\gamma_0$ for the three different values of $\sigma$.
(Fig.~\ref{fig:max_mu_star}). The star graph reproduces the same
qualitative trend found on the chain (see Fig.~\ref{fig:im_mu}): as the
gain-loss profile becomes more localized (i.e., as $\sigma$ decreases,
recall the convention of Eq.~\eqref{Gaussian}), the $\mathcal{PT}$-breaking
threshold shifts to larger values of $\gamma_0$, since fewer sites near
the vertex participate in the non-Hermitian interaction.

\subsection{Anticontinuum limit ($\kappa_1=0$): the single-vertex soliton}
\label{subsec:Anticontinuum1}

Having established the spectral properties of the star graph in previous subsection, we now turn to the nonlinear regime
$g\neq 0$ and construct soliton solutions on the branched
lattice, following the same anticontinuum continuation strategy as for the
chain (Subsection~\ref{subsec:existence}).

Setting $\kappa_1=0$ decouples the lattice into isolated units. On each edge,
away from the vertex, sites obey the uncoupled dimer equation and we take the
trivial solution $\Omega_n=0$ for $n\neq0$. At the vertex, Eq.~\eqref{uv1}
reduces to
\begin{align}
\begin{split}
 &   (\mu-i\gamma_0)u_{-j,0}+\kappa_0(v_{1,0}+v_{2,0})=g|u_{-j,0}|^2u_{-j,0},
\\
&   (\mu+i\gamma_0)v_{j,0}+\kappa_0(u_{-1,0}+u_{-2,0})=g|v_{j,0}|^2v_{j,0},
\label{eq:vertex-anticont}
\end{split}
\end{align}
for $j=1,2$.

Imposing the $\mathcal{PT}$-symmetry reduction $u_{-j,0}=\bar v_{j,0}$ and
setting $p=u_{-1,0}=\bar v_{1,0}$, $q=u_{-2,0}=\bar v_{2,0}$, only two
independent equations remain:
\begin{equation}
(\mu-i\gamma_0)p+\kappa_0(\bar p+\bar q)=g|p|^2p,
\qquad
(\mu-i\gamma_0)q+\kappa_0(\bar p+\bar q)=g|q|^2q.
\label{eq:pq}
\end{equation}

Subtracting the two equations in \eqref{eq:pq} splits the problem into an
antisymmetric branch ($p=-q$) and a symmetric branch ($p=q$). 
The more
general case $p\neq\pm q$, which gives rise to vortex-type solitons
carrying a nonzero winding number, is treated separately in
Section~\ref{sec:vortex}.

\paragraph{Antisymmetric branch ($p=-q$)}
In this branch the coupling term cancels, and Eq.~\eqref{eq:pq} reduces to
\begin{equation}
(\mu-i\gamma_0)p=g|p|^2p.
\label{eq:antisym-eq}
\end{equation}
Since the right-hand side is real, a nontrivial solution ($p\neq0$) requires
$\gamma_0=0$ and for any $\gamma_0\neq0$ only the trivial solution $p=0$ exists.
The antisymmetric branch cannot bifurcate once gain-loss is present: the
two gain sites drive the hub in exactly opposite phase, so their net
coupling to the loss sites cancels.
At $\gamma_0=0$, Eq.~\eqref{eq:antisym-eq} becomes $\mu p=g|p|^2p$, giving the
explicit solution
\begin{equation}
p=\sqrt{\frac{\mu}{g}}\,e^{i\theta},\quad \theta\in[0,2\pi),
\label{eq:antisym-solution}
\end{equation}
which exists provided $\mu/g>0$. The phase $\theta$ is undetermined because
at $\gamma_0=0$ the system possesses a continuous gauge symmetry
$p\mapsto pe^{i\theta}$. Jacobian matrix of this branch
is singular which will be shown later, and the branch cannot be continued
by the standard implicit-function-theorem argument used below. We therefore
disregard this branch in what follows and consider the
$\mathcal{PT}$-symmetric case $\gamma_0\neq0$.

\paragraph{Symmetric branch ($p=q$)}
For $p=q$, Eq.~\eqref{eq:pq} becomes a single dimer-type equation with
doubled effective coupling,
\begin{equation}
(\mu-i\gamma_0)p+2\kappa_0\bar p=g|p|^2p,
\label{eq:symmetric-dimer}
\end{equation}
which can be solved by the ansatz $p=Ae^{i\phi}$. Substituting
and separating real and imaginary parts gives the exact analytical solution as
\begin{equation}
p_{\pm} = A_{\pm} e^{i\phi_{\pm}}, \;\;
A_{\pm}^2=\frac{\mu\pm\sqrt{4\kappa_0^2-\gamma_0^2}}{g}, \;\;
\sin(2\phi_\pm)=-\frac{\gamma_0}{2\kappa_0}.
\label{eq:vertex-solution}
\end{equation}

Two independent conditions must hold for
this branch 
to exist as a nontrivial, real-amplitude solution:
\begin{enumerate}
\item[(i)] The phase equation $\sin(2\phi_\pm)=-\gamma_0/(2\kappa_0)$ requires
\begin{equation}
\gamma_0<2\kappa_0.
\label{eq:vertex-PT-condition}
\end{equation}
This guarantees $\cos(2\phi_\pm)=\pm\sqrt{1-\gamma_0^2/4\kappa_0^2}$ is real.
\item[(ii)] Defining $\mu_\pm:=\mp\sqrt{4\kappa_0^2-\gamma_0^2}$, positivity of the amplitude $A_{\pm}$ requires
\begin{equation}
\mu>\mu_\pm.
\label{eq:vertex-existence-mu}
\end{equation}
\end{enumerate}
Conditions \eqref{eq:vertex-PT-condition} and \eqref{eq:vertex-existence-mu}
together define the region of parameters $(\mu,\gamma_0)$ in which the
symmetric solution $p_\pm$ exists.

\subsection{Continuation to $\kappa_1>0$: invertibility of the Jacobian}

Having identified the antisymmetric and symmetric branches at $\kappa_1=0$,
we now examine, by an exact, analytical argument, which of these branches can be analytically continued to
$\kappa_1>0$.
To analyze the analytical continuation of stationary solutions (from $\kappa_1=0$ to $\kappa_1>0$) and to apply the implicit function theorem, it is convenient to rewrite the complex amplitudes in terms of their real and imaginary parts,
\[
p=a+ib,\quad q=c+id,
\]
where $a,b,c,$ and $d$ are real-valued variables. Substituting these expressions into the stationary equations yields the following system of algebraic equations:
\begin{align}
\begin{split}
& \mu a+\gamma_0 b+\kappa_0(a+c)
=
g(a^2+b^2)a, \\
& \mu b-\gamma_0 a-\kappa_0(b+d)
=
g(a^2+b^2)b, \\
& \mu c+\gamma_0 d+\kappa_0(a+c)
=
g(c^2+d^2)c, \\
& \mu d-\gamma_0 c-\kappa_0(b+d)
=
g(c^2+d^2)d.
\label{eq:real4}
\end{split}
\end{align}

The Jacobian matrix of the nonlinear system
\eqref{eq:real4}
plays a central role in the local bifurcation analysis.
In particular, the invertibility of the Jacobian guarantees,
via the implicit function theorem,
the local continuation of stationary solutions.
The Jacobian matrix can be written in the simplified form
\begin{align}
\begin{split}
& J=\begin{pmatrix}
    J_{11} & J_{12} \\ J_{21} & J_{22}
\end{pmatrix}, \quad
J_{12}=J_{21}= \kappa_0 \sigma_3,
\\
& J_{11} = \begin{pmatrix}
-2ga^{2}-\gamma_0\dfrac{b}{a}-\kappa_0\dfrac{c}{a}
& \gamma_0-2gab \\
-\gamma_0-2gab
&
-2gb^{2}+\gamma_0\dfrac{a}{b}
+\kappa_0\dfrac{d}{b}
\end{pmatrix}, \\
& J_{22} = \begin{pmatrix}
-2gc^{2}-\gamma_0\dfrac{d}{c}
-\kappa_0\dfrac{a}{c}
&
\gamma_0-2gcd \\
-\gamma_0-2gcd
&
-2gd^{2}
+\gamma_0\dfrac{c}{d}
+\kappa_0\dfrac{b}{d}
\end{pmatrix},
\end{split}
\label{eq:Jacobian}
\end{align}
where $\sigma_3$ is the Pauli matrix.

The invertibility of the Jacobian matrix determines the parameter region where the stationary solutions can be uniquely continued and is therefore essential for the existence analysis.

\paragraph{Singularity of the Jacobian on the antisymmetric branch}
On the antisymmetric branch we substitute $c=-a$, $d=-b$ (i.e., $q=-p$) into
\eqref{eq:Jacobian}. Since then $c/a=d/b=-1$, $d/c=b/a$, $a/c=b/d=-1$, and
$cd=ab$, a direct inspection shows that the two diagonal blocks coincide $J_{11}=J_{22}=:\mathcal{A}$:
\begin{equation}
\mathcal{A}=
\begin{pmatrix}
-2ga^{2}-\gamma_0\dfrac{b}{a}+\kappa_0 & \gamma_0-2gab\\[1mm]
-\gamma_0-2gab & -2gb^{2}+\gamma_0\dfrac{a}{b}-\kappa_0
\end{pmatrix},
\label{eq:antisym-A}
\end{equation}
so that the Jacobian acquires the block-symmetric form
\begin{equation}
J=\begin{pmatrix}\mathcal{A}&\mathcal{B}\\ \mathcal{B}&\mathcal{A}\end{pmatrix},
\qquad \mathcal{B}=\kappa_0\sigma_3.
\label{eq:antisym-block}
\end{equation}
For any matrix of this form the similarity transformation
$\mathcal{S}=\tfrac{1}{\sqrt{2}}\bigl(\begin{smallmatrix}1&1\\1&-1\end{smallmatrix}\bigr)$,
$\mathcal{S}^{-1}=\mathcal{S}$, block-diagonalizes $J$ without any
commutativity assumption on $\mathcal{A}$ and $\mathcal{B}$:
\begin{equation}
\mathcal{S}J\mathcal{S}^{-1}
=\begin{pmatrix}\mathcal{A}+\mathcal{B}&0\\0&\mathcal{A}-\mathcal{B}\end{pmatrix},
\quad
\det J=\det(\mathcal{A}+\mathcal{B})\,\det(\mathcal{A}-\mathcal{B}).
\label{eq:det-factorization}
\end{equation}
Recalling that the antisymmetric branch exists only at $\gamma_0=0$, where the
stationary condition reads $\mu=g(a^2+b^2)$, the blocks reduce to
\begin{equation}
\mathcal{A}-\mathcal{B}
=-2g\begin{pmatrix}a^{2}&ab\\ab&b^{2}\end{pmatrix},
\qquad
\mathcal{A}+\mathcal{B}
=\begin{pmatrix}-2ga^{2}+2\kappa_0&-2gab\\-2gab&-2gb^{2}-2\kappa_0\end{pmatrix}.
\label{eq:antisym-blocks-gamma0}
\end{equation}
Since $\det(\mathcal{A}-\mathcal{B})\equiv0$ identically, the factorization
\eqref{eq:det-factorization} gives $\det J\equiv0$ on the entire
antisymmetric branch. Consequently, this branch cannot be continued to
$\kappa_1>0$ via the standard implicit-function-theorem argument, and, being
restricted to the conservative limit $\gamma_0=0$, it is excluded from
further analysis.

\paragraph{Invertibility of the Jacobian on the symmetric branch}
On the symmetric branch we substitute $c=a$, $d=b$ (i.e., $q=p$) into
\eqref{eq:Jacobian}. The two diagonal blocks again coincide,
\begin{equation}
J_{11}=J_{22}=:\mathcal{A}=
\begin{pmatrix}
-2ga^{2}-\gamma_0\dfrac{b}{a}-\kappa_0 & \gamma_0-2gab\\[1mm]
-\gamma_0-2gab & -2gb^{2}+\gamma_0\dfrac{a}{b}+\kappa_0
\end{pmatrix},
\label{eq:sym-A}
\end{equation}
so that $J$ has the same block-symmetric structure as in
\eqref{eq:antisym-block}, and the factorization
\eqref{eq:det-factorization} applies:
$\det J=\det(\mathcal{A}+\mathcal{B})\,\det(\mathcal{A}-\mathcal{B})$
with $\mathcal{B}=\kappa_0\sigma_3$.

Evaluating both factors on the solution \eqref{eq:vertex-solution}, i.e.,
substituting $a=A_\pm\cos\phi_\pm$, $b=A_\pm\sin\phi_\pm$ and using
$\sin(2\phi_\pm)=-\gamma_0/(2\kappa_0)$,
$\cos(2\phi_\pm)=\pm\sqrt{4\kappa_0^{2}-\gamma_0^{2}}/(2\kappa_0)$, a direct
computation gives the closed-form expressions
\begin{align}
\det(\mathcal{A}+\mathcal{B})|_{(\pm)}
&=\pm\,4g\,A_{\pm}^{2}\sqrt{4\kappa_0^{2}-\gamma_0^{2}},
\label{eq:det-plus}\\[1mm]
\det(\mathcal{A}-\mathcal{B})|_{(\pm)}
&=4\kappa_0^{2}\pm2g\,A_{\pm}^{2}\sqrt{4\kappa_0^{2}-\gamma_0^{2}}.
\label{eq:det-minus}
\end{align}

The factor \eqref{eq:det-plus} is nonzero whenever the solution exists as a
genuine nonlinear mode, i.e., $A_\pm\neq0$ and $\gamma_0<2\kappa_0$; it
imposes no additional restriction beyond the existence conditions
\eqref{eq:vertex-PT-condition} and \eqref{eq:vertex-existence-mu}.

The factor \eqref{eq:det-minus} distinguishes the two branches. For
$(+)$ it is a sum of two positive terms (for $g>0$), hence
\begin{equation}
\det J\big|_{(+)}\neq0
\qquad\text{for all }\ \gamma_0<2\kappa_0,\ \ \mu>\mu_+,
\label{eq:gamma-plus-invertible}
\end{equation}
so the branch $(+)$ can always be continued to $\kappa_1>0$ by the
implicit function theorem. For branch $(-)$, however,
\eqref{eq:det-minus} vanishes when
$gA_{-}^{2}=2\kappa_0^{2}/\sqrt{4\kappa_0^{2}-\gamma_0^{2}}$, which, using
$gA_-^2=\mu-\mu_-$, translates into the additional condition
\begin{equation}
\mu\ \neq\ \mu_{\mathrm{c}}
:=\frac{6\kappa_0^{2}-\gamma_0^{2}}{\sqrt{4\kappa_0^{2}-\gamma_0^{2}}}.
\label{eq:gamma-minus-condition}
\end{equation}
Hence $(-)$ is continuable for all
$\mu\in(\mu_-,\mu_{\mathrm{c}})\cup(\mu_{\mathrm{c}},\infty)$, i.e., for all
$\mu>\mu_-$ except the single critical value $\mu=\mu_{\mathrm{c}}$.

\subsection{Linear stability of solitons on the branched lattice}
\label{sec:stability-star}

In this subsection, the same manner given in Section~2.3 is used to study the
linear stability of discrete solitons on the four-edge star graph. The
perturbation terms are added to the stationary solution $\Omega_n$ as
\begin{equation}
\Omega_n(t)=\Big(\Omega_n+e^{\lambda t}\,\Psi_n+e^{\bar{\lambda} t}\,\bar{\Phi}_n\Big)\,
e^{-i(\kappa_0+\kappa_1+\mu)t},
\qquad |\Psi_n|,\,|\Phi_n|\ll|\Omega_n|,
\label{eq:star-perturbation}
\end{equation}
where the perturbation blocks $\Psi_n$ and $\Phi_n$ are organized identically
to $\Omega_n$ in \eqref{eq:Omega-n}, i.e.,
\begin{equation}
\Psi_n=
\begin{cases}
\big(\psi^{u}_{1,n},\ \psi^{v}_{1,n},\ \psi^{u}_{2,n},\ \psi^{v}_{2,n}\big)^{T}, & n\ge 1,\\[4pt]
\big(\psi^{u}_{-1,0},\ \psi^{v}_{1,0},\ \psi^{u}_{-2,0},\ \psi^{v}_{2,0}\big)^{T}, & n=0,\\[4pt]
\big(\psi^{u}_{-1,n},\ \psi^{v}_{-1,n},\ \psi^{u}_{-2,n},\ \psi^{v}_{-2,n}\big)^{T}, & n\le -1,
\end{cases}
\label{eq:Psi-def}
\end{equation}
and $\Phi_n$ is defined in exactly the same way.

Substituting \eqref{eq:star-perturbation} into the NLDS equation on the graph,
and linearizing with respect to $\Psi_n$ and $\Phi_n$ 
we obtain the
eigenvalue problem written in a cell as
\begin{multline}
\begin{pmatrix}\mathcal{T}_-&0\\[1mm]0&-\mathcal{T}_-\end{pmatrix}
\begin{pmatrix}\Psi_{n-1}\\[1mm]\Phi_{n-1}\end{pmatrix}
+
\begin{pmatrix}\mathcal{L}_n&\mathcal{G}_n\\[1mm]-\bar{\mathcal{G}}_n&-\bar{\mathcal{L}}_n\end{pmatrix}
\begin{pmatrix}\Psi_n\\[1mm]\Phi_n\end{pmatrix}\\
+
\begin{pmatrix}\mathcal{T}_+&0\\[1mm]0&-\mathcal{T}_+\end{pmatrix}
\begin{pmatrix}\Psi_{n+1}\\[1mm]\Phi_{n+1}\end{pmatrix}
=i\lambda
\begin{pmatrix}\Psi_n\\[1mm]\Phi_n\end{pmatrix},
\;\;
n\in\mathbb{Z},
\label{eq:star-stability}
\end{multline}
where
blocks are expressed through the stationary solution $\Omega_n$ as
\begin{equation}
\mathcal{L}_n=[2g|\Omega_n|^{2}-\mu] I_4+\mathcal{D}_n,
\qquad
\mathcal{G}_n=g\Omega_n^{2} I_4,
\label{eq:LnGn-star}
\end{equation}
and $\mathcal{T}_\pm$, $\mathcal{D}_n$ (including the vertex block
$\mathcal{D}_0$) are given in Eqs.~\eqref{eq:bulk-4x4} and \eqref{D0}.

Note that the vertex modifies the stability problem only through the block
$\mathcal{D}_0$ contained in $\mathcal{L}_0$, the block
$\mathcal{G}_0$ remains diagonal because the nonlinearity acts on-site. If the
coupling $-\kappa_0\sigma_1$ is removed from $\mathcal{D}_0$, the system
\eqref{eq:star-stability} decouples into two independent copies of the chain
problem \eqref{stability}, which serves as a consistency check.

\paragraph{Linear stability in the anticontinuum limit}

At $\kappa_1=0$ the vertex stability matrix takes the block form
\begin{equation}
M=\begin{pmatrix} \mathcal{L}_0 & \mathcal{G}_0 \\ -\mathcal{\bar G}_0 & -\mathcal{\bar L}_0\end{pmatrix},
\qquad
\mathcal{L}_0=\begin{pmatrix}\widetilde A&\widetilde B\\\widetilde B&\widetilde A\end{pmatrix},
\qquad
\mathcal{G}_0=\begin{pmatrix}\widetilde C&0\\0&\widetilde C\end{pmatrix},
\label{eq:M-ABC}
\end{equation}
where, for the symmetric branch $p=q$, and reusing the $2\times2$ block
$D_0$ from \eqref{DT} evaluated at $n=0$,
\begin{equation}
\widetilde A=[2g|p|^2 -\mu] I_2+D_0,
\qquad \widetilde B=-\kappa_0\sigma_1,
\qquad \widetilde C=g \;\text{diag}(p^2, \bar p^2).
\label{eq:ABC-def}
\end{equation}

Following the channel-diagonalization used for the existence Jacobian, let
$S=\tfrac{1}{\sqrt2}\big(\begin{smallmatrix}I&I\\I&-I\end{smallmatrix}\big)$
and $Q=\mathrm{diag}(S,S^*)$, since $S$ is real, $S^*=S$ and $S^{-1}=S$.
Direct computation gives
\begin{equation}
Q^{-1}MQ\sim M_+\oplus M_-,
\qquad
M_\pm=\begin{pmatrix}\widetilde A\pm\widetilde B&\widetilde C\\-\widetilde C^*&-(\widetilde A\pm\widetilde B)^*\end{pmatrix}.
\label{eq:M-blockdiag}
\end{equation}

\begin{figure}[t!]
    \centering
    \includegraphics[width=0.99\linewidth]{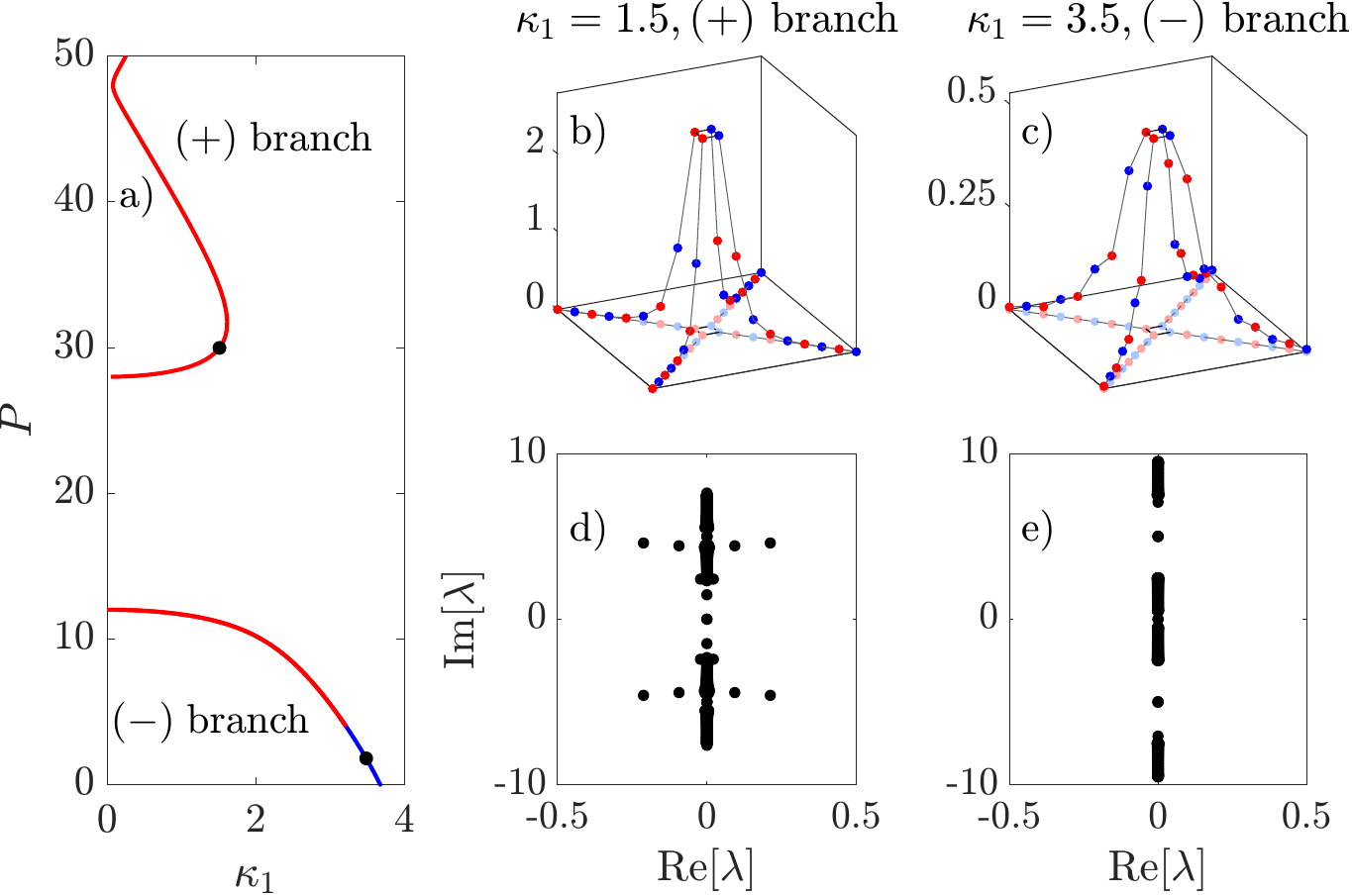}
    \caption{Continuation of the single-vertex solitons to $\kappa_1>0$.
(a) Power $P$ versus $\kappa_1$ for the $(+)$ and $(-)$ branches, for $\kappa_0=1$,
$\mu=5$, $\gamma_0=0.1, \; \sigma=10$. Blue segments denote
stable and red segments unstable solutions. Black dots mark the
representative points examined in panels (b)--(e). (b),(c) Soliton
amplitude profiles at $\kappa_1=1.5$ for the $(+)$ branch and
$\kappa_1=3.5$ for the $(-)$ branch, respectively, with red dots
corresponding to $|u_{\pm j, n}|$ and blue dots to $|v_{\pm j, n}|$.
(d),(e) Corresponding linear stability spectra $\lambda$ in the complex
plane for the $(+)$ and $(-)$ branches, respectively.}
\label{fig:sol_star_full}
\end{figure}

The spectrum of $M$ splits into two independent $4\times4$ eigenvalue
problems, $\det(M_\pm-i\lambda I_4)=0$. Each is identical in structure to the
single-dimer stability matrix of Ref.~\cite{Konotop_2012}, with coupling
$\kappa_0$ replaced by $2\kappa_0$ (block $M_+$) and by $0$ (block $M_-$).

Using the solutions in \eqref{eq:vertex-solution}, the characteristic polynomials reduce to
\begin{align}
&\det(M_+-i\lambda I_4)=\lambda^2\Big[\lambda^2+8\mu_-\,
\Big(\mu_-\pm\frac{\mu}{2}\Big)\Big],
\label{eq:Mplus-poly} \\
& \det(M_--i\lambda I_4)=\lambda^4+4\Big[2\mu_-^2 
-2\kappa_0^2 \pm \mu\mu_-\Big]\lambda^2+4\Big[2\kappa_0^2 + \mu_-^2 \pm \mu\mu_- \Big]^2,
\label{eq:Mminus-poly}
\end{align}
where $\mu_- = \sqrt{4\kappa_0^2-\gamma_0^2}$.
The double root of \eqref{eq:Mplus-poly} is $\lambda=0$, while the remaining two exact roots are purely imaginary
\begin{equation}
\lambda=\pm i\sqrt{8\mu_-\,
\Big(\mu_-\pm \frac{\mu}{2}\Big)}.
\label{eq:lambda-plus}
\end{equation}
Solving \eqref{eq:Mminus-poly} gives the closed form
\begin{equation}
\lambda=\pm\gamma_0\pm \sqrt{\pm 2 \mu\mu_- - 3\mu_-^2}.
\label{eq:lambda-minus}
\end{equation}
Eq.~\eqref{eq:lambda-minus} shows that
$\mathrm{Re}(\lambda)=\pm\gamma_0\neq0$ whenever $\gamma_0\neq0$,
independently of $\mu$, $\kappa_0$, and branch sign (the sign inside the square root). Since the full
$8$-dimensional perturbation space is the direct sum of the two sectors in
\eqref{eq:M-blockdiag}, both the $(+)$ and $(-)$ single-vertex solitons are
unstable throughout the entire anticontinuum limit.

This reflects broken local $\mathcal{PT}$-symmetry at the isolated vertex:
the block $M_-$ eliminates the $\kappa_0$ coupling, and the gain-loss
sites decouple with eigenvalues $\pm\gamma_0$ for any $\gamma_0\neq0$. Broken local $\mathcal{PT}$-symmetry
does not preclude stability of the full lattice, since increasing the
coupling coefficient $\kappa_1$ can restore global $\mathcal{PT}$-symmetry
and stabilize the soliton (see Ref.~\cite{Konotop_2012}). Motivated by this analogy, we obtain numerical results for both
$(\pm)$ branches to $\kappa_1>0$ and search stability
windows at larger $\kappa_1$.

As $\kappa_1$ increases, the $(-)$ branch acquires a finite stability
window at larger $\kappa_1$ (blue segment in panel (a) of
Fig.~\ref{fig:sol_star_full}), in which the antisymmetric instability
identified in \eqref{eq:lambda-minus} is suppressed through hybridization
with the bulk bands.
This is confirmed by the
representative point at $\kappa_1=3.5$ [panel (c), (e)], where the
stability spectrum is confined to the imaginary axis. The $(+)$ branch, by
contrast, remains unstable throughout the range of $\kappa_1$ explored
here, as illustrated at $\kappa_1=1.5$ [panel (b), (d)], where an isolated
eigenvalue quartet appears off the imaginary axis. This asymmetry between
the two branches motivates a closer look, in the following section, at a
qualitatively different class of vertex states.

\section{Vortex solitons on the star graph}
\label{sec:vortex}

\subsection{Existence and topological characterization}

The single-vertex solitons constructed in
Subsection~\ref{subsec:Anticontinuum1}, restricted to the symmetric ($p=q$)
and antisymmetric ($p=-q$) branches, exhaust only a special subset of the
stationary solutions at the vertex. Moreover, the stability analysis of
Subsection~\ref{sec:stability-star} revealed a marked asymmetry between them,
with only the $(-)$ branch acquiring a stability window at larger
$\kappa_1$. This motivates a qualitatively different class of vertex
states, in which the amplitudes at the two incoming edges differ in
magnitude, $|p|\neq|q|$, but wind in phase around the vertex -- the
discrete analogue of a vortex soliton, characterized by an integer winding
number. We show that such a state exists at the vertex of the
$\mathcal{PT}$-symmetric star graph, and examine its existence, topological
characterization, and linear stability.

Eq.~\eqref{eq:pq} was previously analyzed only along the two special
branches $p=q$ and $p=-q$ (Subsection~\ref{subsec:Anticontinuum1}). Subtracting the two equations in
\eqref{eq:pq},
\begin{equation}
(\mu-i\gamma_0)(p-q)=g\big(|p|^2p-|q|^2q\big),
\label{eq:pq_diff}
\end{equation}
does not by itself restrict solutions to $p=\pm q$: \eqref{eq:pq_diff}
is merely a necessary consequence of \eqref{eq:pq}, and the full system
admits additional solutions, in particular with
$|p|\neq|q|$. We now construct this family explicitly and show that it
carries a nonzero discrete winding number around the vertex.

\paragraph{Exact solution at $\gamma_0=0$}

At $\gamma_0=0$, both components of $p,q$ may be taken real. Writing
$p=-P$, $q=Q$ with $P,Q>0$, Eq.~\eqref{eq:pq} reduces to
\begin{equation}
-\mu P+\kappa_0(Q-P)=-gP^3,\qquad
\mu Q+\kappa_0(Q-P)=gQ^3,
\label{eq:vortex-real-eqs}
\end{equation}
whose sum and difference give
\begin{equation}
P^2+Q^2=\frac{\mu+\kappa_0}{g},
\qquad
P^2 Q^2=\Big(\frac{\kappa_0}{g}\Big)^2.
\label{eq:vortex-sum-prod}
\end{equation}
Thus $P^2$ and $Q^2$ are the two roots of the quadratic equation that gives the exact solutions as
\begin{equation}
P^2,\,Q^2=\frac{(\mu+\kappa_0)\mp\sqrt{(\mu+\kappa_0)^2-4\kappa_0^2}}{2g},
\label{eq:vortex-AB}
\end{equation}
real and positive provided $\mu>\kappa_0$. 
The full vertex state is
\begin{equation}
u_{-1,0}=v_{1,0}=-P,\qquad u_{-2,0}=v_{2,0}=Q,
\label{eq:vortex-gamma0}
\end{equation}
up to an overall (discrete) phase $p\mapsto -p$, $q\mapsto-q$. A continuous
gauge freedom, present for the antisymmetric branch \eqref{eq:antisym-eq}, is
absent here because \eqref{eq:vortex-real-eqs} is not invariant
under a general phase rotation of $(p,q)$.

\paragraph{Winding number}

Define the discrete winding number of the vertex configuration
$(x_1,x_2,x_3,x_4)=(u_{-1,0},v_{1,0},u_{-2,0},v_{2,0})$ around the
$\kappa_0$-cycle $x_1\!\to\!x_2\!\to\!x_3\!\to\!x_4\!\to\!x_1$ as
\begin{equation}
\ell=\frac{1}{2\pi}\sum_{k=1}^{4}\mathrm{Arg}\Big(\frac{x_{k+1}}{x_k}\Big)\in\mathbb{Z},
\qquad x_5\equiv x_1,
\label{eq:winding-def}
\end{equation}
with each argument taken in $(-\pi,\pi]$. For \eqref{eq:vortex-gamma0},
$(x_1,x_2,x_3,x_4)=(-P,-P,Q,Q)$, and \eqref{eq:winding-def} evaluates to
$\ell=\pm1$ (the sign depending on the branch cut convention at the two
$\pi$-phase jumps), confirming that this state carries nonzero winding number, in contrast to the $p=q$ and $p=-q$ branches, for which
$\ell=0$ identically.

We note that at $\gamma_0=0$ the configuration \eqref{eq:vortex-gamma0} is
purely real and therefore carries zero discrete current,
$\mathcal{J}_{k,k+1}=2\,\mathrm{Im}(\bar x_kx_{k+1})=0$. Here winding number $\ell=\pm1$ reflects a
$\pi$-phase pattern rather than circulation. For $\gamma_0\neq0$,
numerical continuation of \eqref{eq:vortex-gamma0} shows $p,q$ acquire
nonzero imaginary parts, and $\mathcal{J}_{k,k+1}\neq0$, than the vortex becomes a
genuine current-carrying state and still has nonzero winding number.

\paragraph{Continuation in $(\gamma_0,\kappa_1)$}

Writing the real/imaginary parts of \eqref{eq:pq} as a $4$-dimensional
system with $p=a+ib$, $q=c+id$ (and
$\kappa_1$ entering through the bulk coupling once $\Omega_n\neq0$ for
$n\neq0$), the existence Jacobian evaluated at the solution \eqref{eq:vortex-gamma0} ($\gamma_0=0$,
$\kappa_1=0$, $a=-P$, $b=0$, $c=Q$, $d=0$) reads
\begin{equation}
J=
\begin{pmatrix}
-\mu-\kappa_0+3gP^2 & 0 & \kappa_0 & 0\\
0 & \mu-\kappa_0-gP^2 & 0 & -\kappa_0\\
\kappa_0 & 0 & \mu+\kappa_0-3gQ^2 & 0\\
0 & -\kappa_0 & 0 & \mu-\kappa_0-gQ^2
\end{pmatrix}.
\label{eq:vortex-Jacobian}
\end{equation}
Using the relations $gP^2+gQ^2=\mu+\kappa_0$ and $g^2P^2Q^2=\kappa_0^2$
from \eqref{eq:vortex-sum-prod}, direct evaluation of \eqref{eq:vortex-Jacobian}
gives the analytical form
\begin{equation}
\det J=4\kappa_0(\mu-\kappa_0)^2(\mu+3\kappa_0).
\label{eq:vortex-detJ}
\end{equation}
Since $\kappa_0>0$ and $\mu+3\kappa_0>0$ throughout the domain of interest,
\eqref{eq:vortex-detJ} vanishes only at $\mu=\kappa_0$. This is precisely the
boundary of the existence condition $\mu>\kappa_0$ already established
in Eq.~\eqref{eq:vortex-AB}. Thus $\det J\neq0$ everywhere the
vortex solution exists.

By the implicit function theorem applied jointly in
$(\gamma_0,\kappa_1)$, the vortex solution \eqref{eq:vortex-gamma0}
therefore extends to a unique, smooth family $\Omega(\gamma_0,\kappa_1)$
on an open neighborhood of $(0,0)$ in parameter space.

\begin{figure}[t!]
    \centering
    \includegraphics[width=0.99\linewidth]{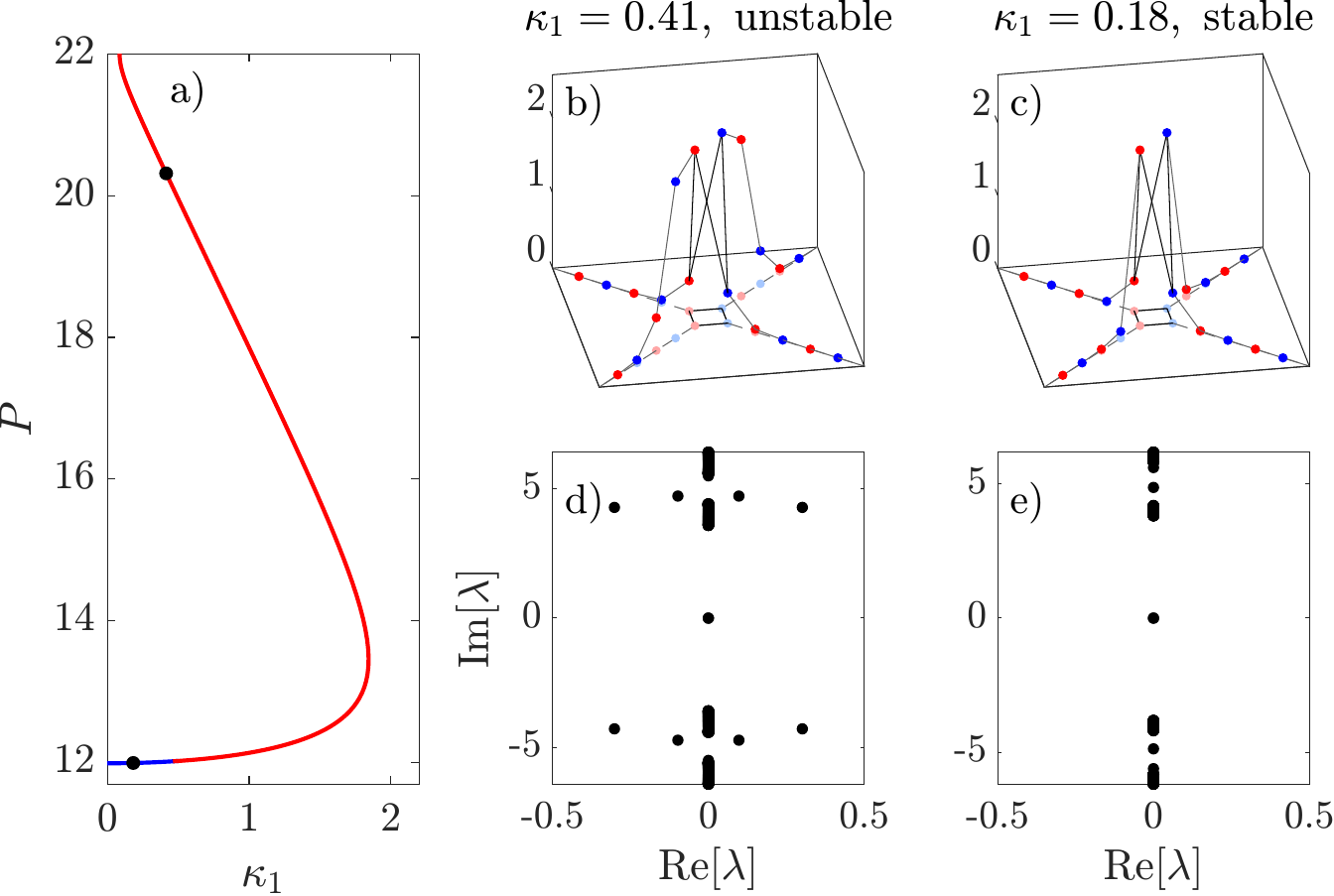}
    \caption{Continuation of the vortex soliton for the parameters $\kappa_0=1,\;\gamma_0=0.1, \; \sigma=10$ and $\mu=5$. (a) Power $P$ versus
    $\kappa_1$; blue segment denotes stable and red unstable solutions.
    Black dots mark the representative points examined in panels
    (b)--(e). (b),(c) Vortex amplitude profiles at $\kappa_1=0.41$
    (unstable) and $\kappa_1=0.18$ (stable), respectively, with red and
    blue dots denoting $|u_{\pm j,\pm n}|$ and $|v_{\pm j,\pm n}|$. (d),(e)
    Corresponding linear stability spectra $\lambda$ in the complex
    plane.}
    \label{fig:vortex_soliton}
\end{figure}

\subsection{Linear stability at $\kappa_1=0$}

The stability matrix $M$ is evaluated directly on the vortex solution
\eqref{eq:vortex-gamma0}. Since $|p|\neq |q|$, the channel-swap
block-diagonalization used in Subsection~\ref{sec:stability-star} does not apply here,  
so the full $8\times8$ eigenvalue problem
$\det(M-i\lambda I_8)=0$ must be solved directly. Using the relations
\eqref{eq:vortex-sum-prod} to eliminate $P,Q$ in favor of $\mu,\kappa_0$,
the characteristic polynomial factors exactly as
\begin{equation}
D(\lambda)=\lambda^2\Big[\lambda^2+(\mu-\kappa_0)(\mu+3\kappa_0)\Big]
\Big[\lambda^4+(\mu^2+2\kappa_0\mu+5\kappa_0^2)\lambda^2
+4\kappa_0\big(\mu^3-7\kappa_0^2\mu+6\kappa_0^3\big)\Big],
\label{eq:vortex-Dlambda}
\end{equation}
where the double root $\lambda=0$ reflects gauge invariance. A direct
sign analysis of \eqref{eq:vortex-Dlambda} on the existence domain
$\mu>\kappa_0$ \eqref{eq:vortex-AB} shows that all eight eigenvalues are
purely imaginary if and only if
\begin{equation}
\mu\in(2\kappa_0,\,\mu_2)\ \cup\ (\mu_3,\,\infty),
\;\;
\mu_2=\kappa_0\big(3+2\sqrt{5-2\sqrt3}\big),
\;\;
\mu_3=\kappa_0\big(3+2\sqrt{5+2\sqrt3}\big),
\label{eq:vortex-stable}
\end{equation}
i.e.\ $\mu_2\approx5.479\,\kappa_0$ and $\mu_3\approx8.819\,\kappa_0$.
Outside these windows the vortex is linearly unstable. A real eigenvalue
pair appears for $\kappa_0<\mu<2\kappa_0$, and a complex quartet
appears for $\mu_2<\mu<\mu_3$. This is in marked contrast to the $p=q$
branch, for which \eqref{eq:lambda-minus} gives
$\mathrm{Re}(\lambda)=\pm\gamma_0\neq0$ for any $\gamma_0\neq0$. The
vortex is stable over two disjoint, nonempty windows in $\mu$ where the
standard single-vertex soliton never is.

\begin{figure}[t!]
    \centering
    \includegraphics[width=0.49\linewidth]{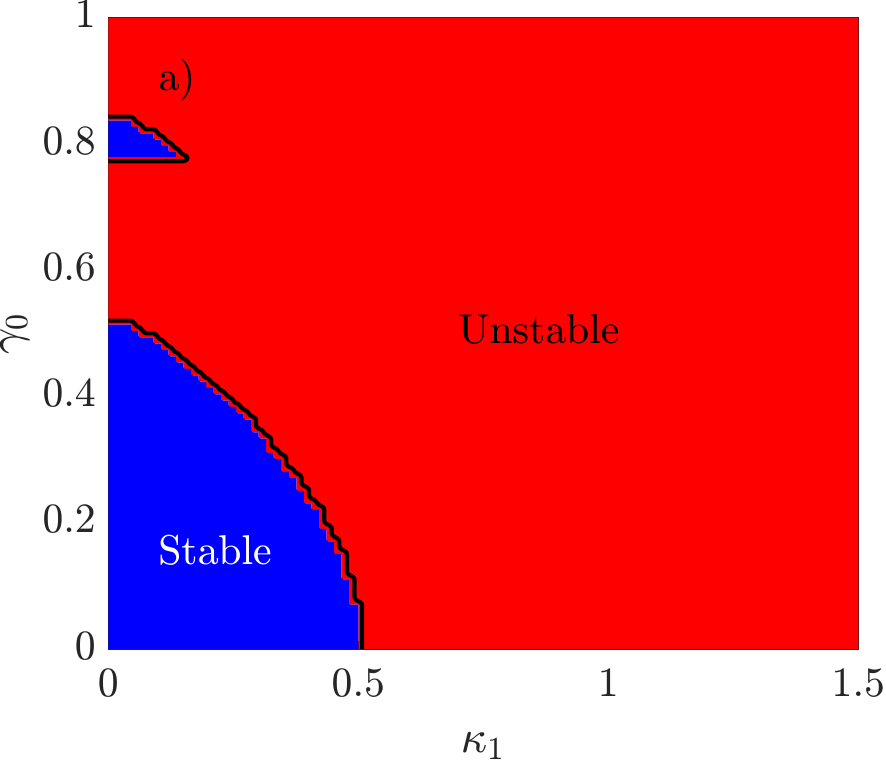}
    \includegraphics[width=0.49\linewidth]{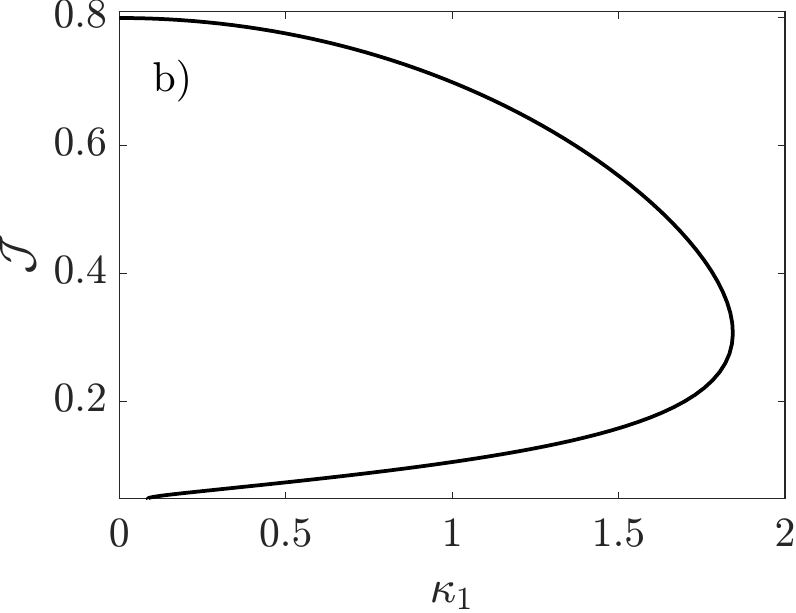}
    \caption{(a) Linear stability of the vortex soliton
    \eqref{eq:vortex-gamma0} on the plane $(\kappa_1,\gamma_0)$, for
    $\mu=5$, $\kappa_0=1$, $\sigma=10$. Blue: linearly stable
    ($\max[\mathrm{Re}(\lambda)]\approx0$); red: linearly unstable. 
    A small,
    disconnected stability island appears near $\gamma_0\approx0.8$ at
    small $\kappa_1$. The black curve marks the corresponding stability
    boundary at $N=10$. (b) Vertex current $\mathcal{J}=\sum_{k=1}^4\mathcal{J}_{k,k+1}$ as a function of $\kappa_1$, for $\gamma_0=0.1$, up to
    the fold at $\kappa_1\approx1.84$.}
    \label{fig:stability_current}
\end{figure}

Fig.~\ref{fig:vortex_soliton} shows the continuation of the vortex
soliton \eqref{eq:vortex-gamma0} in $\kappa_1$, restricted to the primary
branch up to the fold at $\kappa_1\approx1.84$. The vortex is stable only
in a narrow window near $\kappa_1\approx0.2$ [panel (c)], while it becomes
unstable, with an eigenvalue quartet off the imaginary axis, for larger
$\kappa_1$ [panel (b)]. Throughout this range the winding number remains
fixed at $\ell=-1$, and the vertex current $\mathcal{J}_{k,k+1}$ varies
smoothly (see Fig.~\ref{fig:stability_current}b), confirming that the vortex retains a
constant winding number.

The vortex soliton profile (b and c panels of Fig.~\ref{fig:vortex_soliton}) shows that only a few sites in the immediate
vicinity of the vertex are appreciably excited, the amplitude decaying
rapidly along each edge, the truncation length is therefore expected to
have little effect on the results. We confirmed this by repeating the
continuation and stability analysis with $N=10$ (a five-fold reduction
from $N=50$): over the full range, the power
$P(\kappa_1)$ differs by an amount of order $10^{-8}$ and the stability
margin $\max\,\mathrm{Re}(\lambda)$ by on the order of $10^{-3}$, the
stability transition near $\kappa_1\approx0.469$ shifts by less than
$10^{-3}$, and the winding number $\ell=-1$ is reproduced identically.

Fig.~\ref{fig:stability_current}a shows the linear stability of the
vortex soliton \eqref{eq:vortex-gamma0} on the $(\kappa_1,\gamma_0)$
plane, obtained by numerically continuing the solution away from
$\gamma_0=0$ and $\kappa_1=0$. The dominant stable region occupies small
$\kappa_1$ and small-to-moderate $\gamma_0$; as $\kappa_1$ increases
toward $\kappa_1\approx0.5$, the range of $\gamma_0$ supporting a stable
vortex shrinks and vanishes, indicating that increasing the inter-dimer
coupling destabilizes the vortex, in contrast to the single-vertex $(-)$
branch. A second, disconnected stability island appears
near $\gamma_0\approx0.8$ at small $\kappa_1$.

\begin{figure}[t!]
    \centering
    \includegraphics[width=0.6\linewidth]{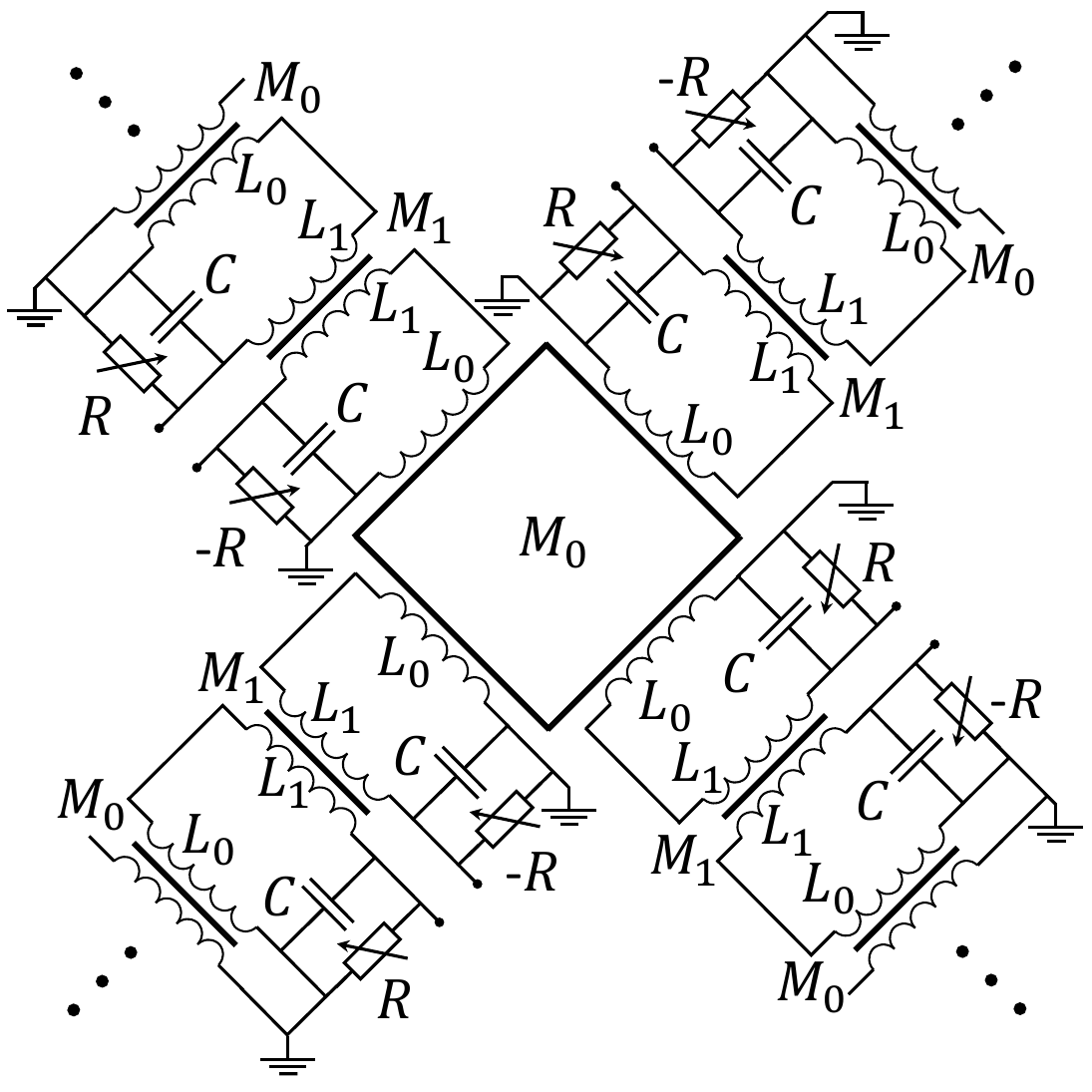}
    \caption{Sketch of a possible electric-circuit realization of the star
graph. Each edge is a chain of parallel $R,\; C,\; L=L_0+L_1$ resonators coupled
by alternating mutual inductances $M_0,M_1$, as in
Ref.~\cite{wang_observation_2020}. The vertex coupling is shown schematically via a
common-core transformer.}
    \label{fig:circuit_star}
\end{figure}

To assess the effect of the truncation length, we additionally
overlay, as a black curve, the corresponding stability boundary obtained
at $N=10$ (see Fig.~\ref{fig:stability_current}a). The two boundaries coincide closely (including around the
isolated stability island near $\gamma_0 \approx 0.8$) with a shift of
order $10^{-2}$ in $\kappa_1$.

Fig.~\ref{fig:stability_current}b shows the corresponding vertex total current
$\mathcal{J}$, for $\gamma_0=0.1$. The
current remains strictly positive and varies smoothly, consistent with
the constant winding number $\ell=-1$.

\section{Sketch for experimental realization}
\label{sketch}

As a possible experimental platform, the star graph considered here could
be realized as a network of coupled electric $LC$ circuits with gain and
loss, following the same strategy used for the simplest
$\mathcal{PT}$-symmetric dimer~\cite{wang_observation_2020}. Each edge is a chain
of parallel $R,\;C$ and $L=L_0+L_1$ resonators coupled by alternating mutual
inductances $M_0,\; M_1$, as sketched in Fig.~\ref{fig:circuit_star}.
We stress that this is only a conceptual sketch of a
possible platform, not a validated realization, for three reasons.
First, the circuit contains only linear $R$, $C$, $L$ elements and thus
has no physical origin for the onsite nonlinearity $g|q_n|^2q_n$; an
explicit nonlinear element (e.g.\ a varactor diode) would be needed.
Second, for simplicity, the sketch depicts the four hub inductors as
sharing a single common core. However, this would also induce unnecessary
direct coupling between the gain-gain sites and between the loss-loss
sites, which is absent in our coupling structure~\eqref{uv1}. A faithful
realization would instead require four independent, edge-by-edge
mutual-inductance elements (or an equivalent common-capacitance coupling
scheme) connecting only the gain--loss pairs. Third, we
have not derived the reduction of the (second-order) mutual-inductive
circuit equations to the first-order NLDS model of Eq.~\eqref{eq:NLDS-V}, nor
the relation between $M_{0}, \; M_1$ and $\kappa_{0},\, \kappa_1$. We leave the
resolution of these points to future engineering-level design.

\section{Conclusions} \label{sec:conclusion}
In this paper we studied discrete $\mathcal{PT}$-symmetric solitons on a four-edge star graph, obtained by coupling four discrete nonlinear Schrödinger chains at a vertex with localized Gaussian gain-loss, via continuation from the anticontinuum limit and linear stability analysis.

Beyond the two conventional vertex branches~\eqref{eq:vertex-solution}, we found a vortex branch \eqref{eq:vortex-AB}, with amplitudes winding in phase around the vertex and carrying a fixed nonzero winding number. It develops a vertex current once gain-loss is switched on, and its stability windows are absent for the solitons in~\eqref{eq:vertex-solution}.

This shows branching is not merely a geometric change to a chain: a multi-edge vertex supports phase-winding states with stability governed jointly by topology, nonlinearity, and $\mathcal{PT}$ symmetry. More broadly, these findings identify branching topology as a new mechanism for engineering localized nonlinear states in non-Hermitian graph-based systems.

The proposed circuit realization offers a route to testing these states experimentally. Extending the construction to other graph topologies and non-Hermitian lattices is a natural next step.

An interesting open direction is whether star graphs with higher-degree vertices, such as a six-edge junction, can support vortex solitons with higher winding numbers $|\ell|>1$, motivated by symmetry-restricted discrete vortices reported in~\cite{GarciaMarch2009}.

\bibliographystyle{unsrt}
\bibliography{references}

@article{exner_quantum_1988,
	title = {Quantum interference on graphs controlled by an external electric field},
	volume = {21},
	issn = {0305-4470},
	url = {https://doi.org/10.1088/0305-4470/21/21/010},
	doi = {10.1088/0305-4470/21/21/010},
	language = {en},
	number = {21},
	journal = {J. Phys. A: Math. Gen.},
	author = {Exner, P. and Seba, P. and Stovicek, P.},
	month = nov,
	year = {1988},
	pages = {4009},
}

@article{kostrykin_kirchhoffs_1999,
	title = {Kirchhoff's rule for quantum wires},
	volume = {32},
	issn = {0305-4470},
	url = {https://doi.org/10.1088/0305-4470/32/4/006},
	doi = {10.1088/0305-4470/32/4/006},
	language = {en},
	number = {4},
	journal = {J. Phys. A: Math. Gen.},
	author = {Kostrykin, V. and Schrader, R.},
	month = jan,
	year = {1999},
	pages = {595},
}

@article{kottos_periodic_1999,
	title = {Periodic {Orbit} {Theory} and {Spectral} {Statistics} for {Quantum} {Graphs}},
	volume = {274},
	issn = {0003-4916},
	url = {https://www.sciencedirect.com/science/article/pii/S0003491699959043},
	doi = {10.1006/aphy.1999.5904},
	number = {1},
	journal = {Annals of Physics},
	author = {Kottos, Tsampikos and Smilansky, Uzy},
	month = may,
	year = {1999},
	pages = {76--124},
}

@book{berkolaiko_introduction_2013,
	title = {Introduction to {Quantum} {Graphs}},
	series = {Mathematical {Surveys} and {Monographs}},
	volume = {186},
	isbn = {978-0-8218-9211-4},
	url = {https://pubs.ams.org/ebooks/surv/186},
	doi = {10.1090/surv/186},
	publisher = {American Mathematical Society},
	author = {Berkolaiko, Gregory and Kuchment, Peter},
	year = {2013},
}

@article{sobirov_integrable_2010,
	title = {Integrable nonlinear {Schrödinger} equation on simple networks: {Connection} formula at vertices},
	volume = {81},
	shorttitle = {Integrable nonlinear {Schrödinger} equation on simple networks},
	url = {https://link.aps.org/doi/10.1103/PhysRevE.81.066602},
	doi = {10.1103/PhysRevE.81.066602},
	number = {6},
	journal = {Phys. Rev. E},
	publisher = {American Physical Society},
	author = {Sobirov, Z. and Matrasulov, D. and Sabirov, K. and Sawada, S. and Nakamura, K.},
	month = jun,
	year = {2010},
	pages = {066602},
}

@article{nakamura_transport_2011,
	title = {Transport in simple networks described by an integrable discrete nonlinear {Schrödinger} equation},
	volume = {84},
	url = {https://link.aps.org/doi/10.1103/PhysRevE.84.026609},
	doi = {10.1103/PhysRevE.84.026609},
	number = {2},
	journal = {Phys. Rev. E},
	publisher = {American Physical Society},
	author = {Nakamura, K. and Sobirov, Z. A. and Matrasulov, D. U. and Sawada, S.},
	month = aug,
	year = {2011},
	pages = {026609},
}

@article{GarciaMarch2009,
  title = {Symmetry, winding number, and topological charge of vortex solitons in discrete-symmetry media},
  author = {Garc\'{\i}a-March, Miguel-\'Angel and Ferrando, Albert and Zacar\'es, Mario and Sahu, Sarira and Ceballos-Herrera, Daniel E.},
  journal = {Phys. Rev. A},
  volume = {79},
  issue = {5},
  pages = {053820},
  numpages = {6},
  year = {2009},
  month = {May},
  publisher = {American Physical Society},
  doi = {10.1103/PhysRevA.79.053820},
  url = {https://link.aps.org/doi/10.1103/PhysRevA.79.053820}
}

@article{Konotop_2012,
	doi = {10.1209/0295-5075/100/56006},
	url = {https://doi.org/10.1209/0295-5075/100/56006},
	year = {2012},
	month = {dec},
	publisher = {EDP Sciences, IOP Publishing and Società Italiana di Fisica},
	volume = {100},
	number = {5},
	pages = {56006},
	author = {Konotop, V. V. and Pelinovsky, D. E. and Zezyulin, D. A.},
	title = {Discrete solitons in $\mathcal{PT}$-symmetric lattices},
	journal = {Europhysics Letters},
}

@article{kevrekidis_nonlinear_2013b,
	title = {Nonlinear {Stationary} {States} in {PT}-{Symmetric} {Lattices}},
	volume = {12},
	url = {https://epubs.siam.org/doi/10.1137/130912694},
	doi = {10.1137/130912694},
	language = {en},
	number = {3},
	journal = {SIAM Journal on Applied Dynamical Systems},
	author = {Kevrekidis, Panayotis G. and Pelinovsky, Dmitry E. and Tyugin, Dmitry Y.},
	year = {2013},
	pages = {1210--1236},
}

@article{li_nonlinear_2012,
	title = {Nonlinear \$ {\textbackslash}mathcal \{{PT}\}\$-symmetric plaquettes},
	volume = {45},
	issn = {1751-8121},
	url = {https://iopscience.iop.org/article/10.1088/1751-8113/45/44/444021},
	doi = {10.1088/1751-8113/45/44/444021},
	number = {44},
	journal = {J. Phys. A: Math. Theor.},
	publisher = {IOP Publishing},
	author = {Li, Kai and Kevrekidis, P. G. and Malomed, Boris A. and Günther, Uwe},
	month = oct,
	year = {2012},
	pages = {444021},
}

@article{li_span_2011,
	title = {{PT}-symmetric oligomers: {Analytical} solutions, linear stability, and nonlinear dynamics},
	volume = {83},
	url = {https://journals.aps.org/pre/abstract/10.1103/PhysRevE.83.066608},
	doi = {10.1103/PhysRevE.83.066608},
	number = {6},
	journal = {Phys. Rev. E},
	publisher = {American Physical Society},
	author = {Li, K. and Kevrekidis, P. G.},
	month = jun,
	year = {2011},
	pages = {066608},
}

@article{zezyulin_stationary_2013,
	title = {Stationary modes and integrals of motion in nonlinear lattices with a \$\{{\textbackslash}mathcal \{{PT}\}\}\$-symmetric linear part},
	volume = {46},
	issn = {1751-8121},
	url = {https://iopscience.iop.org/article/10.1088/1751-8113/46/41/415301},
	doi = {10.1088/1751-8113/46/41/415301},
	number = {41},
	journal = {J. Phys. A: Math. Theor.},
	publisher = {IOP Publishing},
	author = {Zezyulin, D. A. and Konotop, V. V.},
	month = sep,
	year = {2013},
	pages = {415301},
}

@article{zezyulin_nonlinear_2012,
	title = {Nonlinear {Modes} in {Finite}-{Dimensional} \${\textbackslash}mathcal\{{P}\}{\textbackslash}mathcal\{{T}\}\$-{Symmetric} {Systems}},
	volume = {108},
	url = {https://link.aps.org/doi/10.1103/PhysRevLett.108.213906},
	doi = {10.1103/PhysRevLett.108.213906},
	number = {21},
	journal = {Phys. Rev. Lett.},
	publisher = {American Physical Society},
	author = {Zezyulin, D. A. and Konotop, V. V.},
	month = may,
	year = {2012},
	pages = {213906},
}

@article{kevrekidis_nonlinear_2013,
	title = {Nonlinear dynamics in {PT}-symmetric lattices},
	volume = {46},
	issn = {1751-8121},
	url = {https://doi.org/10.1088/1751-8113/46/36/365201},
	doi = {10.1088/1751-8113/46/36/365201},
	number = {36},
	journal = {J. Phys. A: Math. Theor.},
	publisher = {IOP Publishing},
	author = {Kevrekidis, Panayotis G and Pelinovsky, Dmitry E and Tyugin, Dmitry Y},
	month = aug,
	year = {2013},
	pages = {365201},
}

@article{pelinovsky_nonlinear_2014,
	title = {Nonlinear modes in a generalized {\textbackslash}{mathcalPT}-symmetric discrete nonlinear {Schrödinger} equation},
	volume = {47},
	issn = {1751-8121},
	url = {https://doi.org/10.1088/1751-8113/47/8/085204},
	doi = {10.1088/1751-8113/47/8/085204},
	number = {8},
	journal = {J. Phys. A: Math. Theor.},
	publisher = {IOP Publishing},
	author = {Pelinovsky, D E and Zezyulin, D A and Konotop, V V},
	month = feb,
	year = {2014},
	pages = {085204},
}

@article{duanmu_linear_2013,
	title = {Linear and nonlinear parity-time-symmetric oligomers: a dynamical systems analysis},
	volume = {371},
	issn = {1364-503X},
	url = {https://doi.org/10.1098/rsta.2012.0171},
	doi = {10.1098/rsta.2012.0171},
	number = {1989},
	journal = {Phil. Trans. R. Soc. A},
	author = {Duanmu, M. and Li, K. and Horne, R. L. and Kevrekidis, P. G. and Whitaker, N.},
	month = apr,
	year = {2013},
	pages = {20120171},
}

@article{suchkov_solitons_2011,
	title = {Solitons in a chain of parity-time-invariant dimers},
	volume = {84},
	url = {https://link.aps.org/doi/10.1103/PhysRevE.84.046609},
	doi = {10.1103/PhysRevE.84.046609},
	number = {4},
	journal = {Phys. Rev. E},
	publisher = {American Physical Society},
	author = {Suchkov, Sergey V. and Malomed, Boris A. and Dmitriev, Sergey V. and Kivshar, Yuri S.},
	month = oct,
	year = {2011},
	pages = {046609},
}

@article{konotop_nonlinear_2016,
	title = {Nonlinear waves in {PT}-symmetric systems},
	volume = {88},
	url = {https://journals.aps.org/rmp/abstract/10.1103/RevModPhys.88.035002},
	doi = {10.1103/RevModPhys.88.035002},
	number = {3},
	journal = {Rev. Mod. Phys.},
	publisher = {American Physical Society},
	author = {Konotop, Vladimir V. and Yang, Jianke and Zezyulin, Dmitry A.},
	month = jul,
	year = {2016},
	pages = {035002},
}

@article{malomed_discrete_2001,
	title = {Discrete vortex solitons},
	volume = {64},
	url = {https://link.aps.org/doi/10.1103/PhysRevE.64.026601},
	doi = {10.1103/PhysRevE.64.026601},
	number = {2},
	journal = {Phys. Rev. E},
	publisher = {American Physical Society},
	author = {Malomed, B. A. and Kevrekidis, P. G.},
	month = jul,
	year = {2001},
	pages = {026601},
}

@article{neshev_observation_2004,
	title = {Observation of {Discrete} {Vortex} {Solitons} in {Optically} {Induced} {Photonic} {Lattices}},
	volume = {92},
	url = {https://link.aps.org/doi/10.1103/PhysRevLett.92.123903},
	doi = {10.1103/PhysRevLett.92.123903},
	number = {12},
	journal = {Phys. Rev. Lett.},
	publisher = {American Physical Society},
	author = {Neshev, Dragomir N. and Alexander, Tristram J. and Ostrovskaya, Elena A. and Kivshar, Yuri S. and Martin, Hector and Makasyuk, Igor and Chen, Zhigang},
	month = mar,
	year = {2004},
	pages = {123903},
}

@article{leykam_discrete_2013,
	title = {Discrete vortex solitons and parity time symmetry},
	volume = {38},
	issn = {1539-4794},
	url = {https://opg.optica.org/ol/abstract.cfm?uri=ol-38-3-371},
	doi = {10.1364/OL.38.000371},
	number = {3},
	journal = {Opt. Lett.},
	publisher = {Optica Publishing Group},
	author = {Leykam, Daniel and Konotop, Vladimir V. and Desyatnikov, Anton S.},
	month = feb,
	year = {2013},
	pages = {371--373},
}

@article{guo_observation_2009,
	title = {Observation of {P}{T}-{Symmetry} {Breaking} in {Complex} {Optical} {Potentials}},
	volume = {103},
	url = {https://link.aps.org/doi/10.1103/PhysRevLett.103.093902},
	doi = {10.1103/PhysRevLett.103.093902},
	number = {9},
	journal = {Phys. Rev. Lett.},
	publisher = {American Physical Society},
	author = {Guo, A. and Salamo, G. J. and Duchesne, D. and Morandotti, R. and Volatier-Ravat, M. and Aimez, V. and Siviloglou, G. A. and Christodoulides, D. N.},
	month = aug,
	year = {2009},
	pages = {093902},
}

@article{ruter_observation_2010,
	title = {Observation of parity–time symmetry in optics},
	volume = {6},
	issn = {1745-2481},
	url = {https://www.nature.com/articles/nphys1515},
	doi = {10.1038/nphys1515},
	number = {3},
	journal = {Nature Phys},
	publisher = {Nature Publishing Group},
	author = {Rüter, Christian E. and Makris, Konstantinos G. and El-Ganainy, Ramy and Christodoulides, Demetrios N. and Segev, Mordechai and Kip, Detlef},
	month = mar,
	year = {2010},
	pages = {192--195},
}

@article{regensburger_paritytime_2012,
	title = {Parity–time synthetic photonic lattices},
	volume = {488},
	issn = {1476-4687},
	url = {https://www.nature.com/articles/nature11298},
	doi = {10.1038/nature11298},
	number = {7410},
	journal = {Nature},
	publisher = {Nature Publishing Group},
	author = {Regensburger, Alois and Bersch, Christoph and Miri, Mohammad-Ali and Onishchukov, Georgy and Christodoulides, Demetrios N. and Peschel, Ulf},
	month = aug,
	year = {2012},
	pages = {167--171},
}

@article{peng_paritytime-symmetric_2014,
	title = {Parity–time-symmetric whispering-gallery microcavities},
	volume = {10},
	issn = {1745-2481},
	url = {https://www.nature.com/articles/nphys2927},
	doi = {10.1038/nphys2927},
	number = {5},
	journal = {Nature Phys},
	publisher = {Nature Publishing Group},
	author = {Peng, Bo and Özdemir, Şahin Kaya and Lei, Fuchuan and Monifi, Faraz and Gianfreda, Mariagiovanna and Long, Gui Lu and Fan, Shanhui and Nori, Franco and Bender, Carl M. and Yang, Lan},
	month = may,
	year = {2014},
	pages = {394--398},
}

@article{chang_paritytime_2014,
	title = {Parity–time symmetry and variable optical isolation in active–passive-coupled microresonators},
	volume = {8},
	issn = {1749-4893},
	url = {https://www.nature.com/articles/nphoton.2014.133},
	doi = {10.1038/nphoton.2014.133},
	number = {7},
	journal = {Nature Photon},
	publisher = {Nature Publishing Group},
	author = {Chang, Long and Jiang, Xiaoshun and Hua, Shiyue and Yang, Chao and Wen, Jianming and Jiang, Liang and Li, Guanyu and Wang, Guanzhong and Xiao, Min},
	month = jul,
	year = {2014},
	pages = {524--529},
}

@article{wimmer_observation_2015,
	title = {Observation of optical solitons in {PT}-symmetric lattices},
	volume = {6},
	issn = {2041-1723},
	url = {https://www.nature.com/articles/ncomms8782},
	doi = {10.1038/ncomms8782},
	number = {1},
	journal = {Nat Commun},
	publisher = {Nature Publishing Group},
	author = {Wimmer, Martin and Regensburger, Alois and Miri, Mohammad-Ali and Bersch, Christoph and Christodoulides, Demetrios N. and Peschel, Ulf},
	month = jul,
	year = {2015},
	pages = {7782},
}

@article{wang_observation_2020,
	title = {Observation of two {PT} transitions in an electric circuit with balanced gain and loss},
	volume = {74},
	issn = {1434-6079},
	url = {https://doi.org/10.1140/epjd/e2020-10131-7},
	doi = {10.1140/epjd/e2020-10131-7},
	number = {8},
	journal = {The European Physical Journal D},
	author = {Wang, Tishuo and Fang, Jianxiong and Xie, Zhongyi and Dong, Nenghao and Joglekar, Yogesh N. and Wang, Zixin and Li, Jiaming and Luo, Le},
	month = aug,
	year = {2020},
	pages = {167},
}

@article{tabeu_imaginary_2019,
	title = {Imaginary resistor based {Parity}-{Time} symmetry electronics dimers},
	volume = {51},
	issn = {1572-817X},
	url = {https://link.springer.com/article/10.1007/s11082-019-2048-9},
	doi = {10.1007/s11082-019-2048-9},
	number = {10},
	journal = {Opt Quant Electron},
	author = {Tabeu, S. B. and Fotsa-Ngaffo, F. and Kenfack-Jiotsa, A.},
	year = {2019},
	pages = {335},
}

@article{akramov_nonlocal_2022,
	title = {Nonlocal nonlinear {Schrödinger} equation on metric graphs: {A} model for generation and transport of parity-time-symmetric nonlocal solitons in networks},
	volume = {105},
	issn = {2470-0045, 2470-0053},
	url = {https://link.aps.org/doi/10.1103/PhysRevE.105.054205},
	doi = {10.1103/PhysRevE.105.054205},
	language = {en},
	number = {5},
	journal = {Phys. Rev. E},
	author = {Akramov, M. and Sabirov, K. and Matrasulov, D. and Susanto, H. and Usanov, S. and Karpova, O.},
	month = may,
	year = {2022},
	pages = {054205},
}

@article{akramov_discrete_2023,
	title = {Discrete nonlocal nonlinear {Schrödinger} equation on graphs: {Dynamics} of {PT}-symmetric solitons in discrete networks},
	volume = {457},
	issn = {0375-9601},
	url = {https://www.sciencedirect.com/science/article/pii/S0375960122006375},
	doi = {10.1016/j.physleta.2022.128555},
	journal = {Physics Letters A},
	author = {Akramov, M. and Khashimova, F. and Matrasulov, D.},
	month = jan,
	year = {2023},
	pages = {128555},
}

@article{akramov_pt-symmetric_2024,
	title = {{PT}-symmetric solitons in networks: a metric graph based approach},
	volume = {99},
	issn = {1402-4896},
	url = {https://doi.org/10.1088/1402-4896/ad87bb},
	doi = {10.1088/1402-4896/ad87bb},
	language = {en},
	number = {11},
	journal = {Phys. Scr.},
	publisher = {IOP Publishing},
	author = {Akramov, M and Eshchanov, B and Khashimova, F and Matrasulov, D},
	month = oct,
	year = {2024},
	pages = {115279},
}

\end{document}